\documentclass[a4paper,fleqn]{cas-dc}

\usepackage[authoryear,longnamesfirst]{natbib}

\def\tsc#1{\csdef{#1}{\textsc{\lowercase{#1}}\xspace}}
\tsc{WGM}
\tsc{QE}

\usepackage{booktabs,subcaption}

\usepackage{xcolor} 
\usepackage[normalem]{ulem}

\usepackage{xspace}

\begin{document}
\newcommand{\mypar}[1]{\noindent\textbf{#1}}
\newcommand{\mysubpar}[1]{\noindent\emph{\underline{#1}}}
\newcommand{\exescanner}{EXE-scanner\xspace}
\newcommand{\malflip}{MF\xspace}
\newcommand{\goodflip}{GF\xspace}
\newcommand{\todo}[1]{\textcolor{red}{\textbf{[[#1]]}}}
\renewcommand{\sectionautorefname}{Sect.}
\renewcommand{\subsectionautorefname}{Sect.}
\renewcommand{\figureautorefname}{Fig.}
\hyphenation{EXEmples}

\let\WriteBookmarks\relax
\def\floatpagepagefraction{1}
\def\textpagefraction{.001}

\shorttitle{}    

\shortauthors{Kozak et al.}  
\title [mode = title]{Updating Windows Malware Detectors: Balancing Robustness and Regression against Adversarial EXEmples}



%

\author[1]{Matous Kozak}[orcid=0000-0001-8329-7572]
\cormark[1]
\ead{matous.kozak@fit.cvut.cz}



\affiliation[1]{organization={Faculty of Information Technology, Czech Technical University in Prague},
             addressline={Thakurova 9}, 
             city={Prague},
             postcode={16000}, 
             country={Czech Republic}}

\author[2]{Luca Demetrio}


\ead{luca.demetrio@unige.it}



\affiliation[2]{organization={Department of Informatics, Bioengineering, Robotics and Systems Engineering, University of Genova},
             addressline={Viale Causa 13}, 
             city={Genova},
             postcode={16145}, 
             country={Italy}}

\author[2]{Dmitrijs Trizna}
\ead{dmitrijs.trizna@edu.unige.it}

\author[2,3]{Fabio Roli}
\ead{fabio.roli@unige.it}
\affiliation[3]{organization={Department of Electrical and Electronic Engineering,
University of Cagliari},
             addressline={Via Marengo 3}, 
             city={Cagliari},
             postcode={09123}, 
             country={Italy}}




\begin{abstract}
Adversarial EXEmples are carefully-perturbed programs tailored to evade machine learning Windows malware detectors, with an ongoing effort to develop robust models able to address detection effectiveness. 
However, even if robust models can prevent the majority of EXEmples, to maintain predictive power over time, models are fine-tuned to newer threats, leading either to partial updates or time-consuming retraining from scratch. Thus, even if the robustness against adversarial EXEmples is higher, the new models might suffer a regression in performance by misclassifying threats that were previously correctly detected.
For these reasons, we study the trade-off between accuracy and regression when updating Windows malware detectors by proposing \exescanner, a plugin that can be chained to existing detectors to promptly stop EXEmples without causing regression.
We empirically show that previously proposed hardening techniques suffer a regression of accuracy when updating non-robust models, exacerbating the gap when considering low false positives regimes and temporal drifts affecting data.
Also, through \exescanner we gain evidence on the detectability of adversarial EXEmples, showcasing the presence of artifacts left inside while creating them.
Due to its design, \exescanner can be chained to any classifier to obtain the best performance without the need for costly retraining.
To foster reproducibility, we openly release the source code, along with the dataset of adversarial EXEmples based on state-of-the-art perturbation algorithms.
\end{abstract}




\maketitle

\section{Introduction} \label{sec:introduction}
With the pervasive inclusion of machine learning (ML) models in the cybersecurity context, we witness the development and deployment of next-generation Windows antivirus (AV) programs that leverage statistics from data to stop the spread of malware among clients.
In particular, ML models can learn the relationships that exist intrinsically inside data, and they can spot variants of the same malware without creating ad-hoc rules that match a specific sample.
However, these models have been shown to be vulnerable to \emph{adversarial EXEmples}~\citep{demetrio2021adversarial, demetrio2021functionality, kreuk2018deceiving, song2022mab, kolosnjaji2018adversarial}, carefully-crafted Windows malware that preserve malicious logic yet evade detectors through specific manipulations.
These EXEmples are created by messing with the structure of programs through replacement or injection of content that interferes with the decision-making process of models, by also be compliant to the Windows Portable Executable (PE) to keep the functionality intact.
Hence, it is possible to create several variants of EXEmples from a single malware, increasing the likelihood of bypassing a machine learning detector that is unaware of possible mutations.
To face this threat, current ML classifiers need to be updated, either by fine-tuning their performance on the latest data or by retraining them from scratch.
While both practices can increase the overall accuracy, these updates might cause a degradation of performance, as the resulting model starts committing mistakes on previously well-classified programs.
This phenomenon is known as \emph{regression}~\citep{yan2021positive}, and it already affects the update of ML models.
Differently from the computer vision domain where the problem was originally formulated, a regression in the accuracy of Windows malware detectors implies that some malware could be evasive again, while some goodware (benign software) is now rising useless alarms. 
To the best of our knowledge, our paper is among the first in the domain of adversarial EXEmples that studies regression as a metric.

For these reasons, in this paper, we study how practitioners should update machine learning Windows malware detectors to improve the robustness against EXEmples without regression.
We present \emph{\exescanner}, a machine learning model capable of detecting the presence of EXEmples in data that works as a plugin of an existing model to increase its predictive performance while nullifying regression.
\exescanner is trained to recognize most of the proposed types of adversarial EXEmples without the need for replacing the original model.
To do so, we replicate majority of the proposed generators of adversarial EXEmples to create several variants of the same malware, and we publicly release the vast dataset of adversarial EXEmples to foster reproducibility.
We empirically show that the best performance in terms of accuracy and regression is obtained by placing our technique sequentially after the model to harden, and we support this claim by considering different state-of-the-art models and commercial AVs.
This coupling exhibits comparable results to models trained from scratch or fine-tuned to be robust against EXEmples, and it is negligibly impacted by regression.
On the contrary, we show that adversarial training, which is the de-facto standard technique to harden machine learning models by introducing adversarial examples inside the training set~\citep{madry2018towards, lucas2023adversarial}, is more affected by regression. 
Lastly, we analyze why our methodology is able to spot adversarial EXEmples through a-posteriori analysis with SHAP~\cite{NIPS2017_7062}, an explainability method, pointing out that these contain artifacts that can be easily spotted.

Thus, our analysis shows that \exescanner can be easily trained once and attached to any other AV without drawbacks.
To recap, our contributions are the followings: (i) we propose \exescanner, a plugin that can be chained to any AV to improve its robustness against adversarial EXEmples without incurring regression; (ii) we show that \exescanner increases the performances of AVs, matching those that were fine-tuned or retrained from scratch; (iii) we highlight that \exescanner does not suffer from a regression in performance, contrary to state-of-the-art techniques like adversarial training; and (iv) we publicly release the dataset of all the generated adversarial EXEmples we used for our experiments.

\section{Background} \label{sec:background}
Before describing the internals of \exescanner and how we train it, we first introduce the basic concepts needed to better understand our proposal.

\subsection{Malware Detection and Regression}
\label{sec:malware_detection}
Before we introduce our methodology, it is necessary to understand how data in this domain is shaped, by introducing the way Windows programs are stored as files.
Hence, on top of this structure, it is possible to extract meaningful features used to fit machine learning models on such data to discriminate legitimate and harmful programs.

\begin{figure}[t!]
    \centering
    \includegraphics[width=0.45\textwidth]{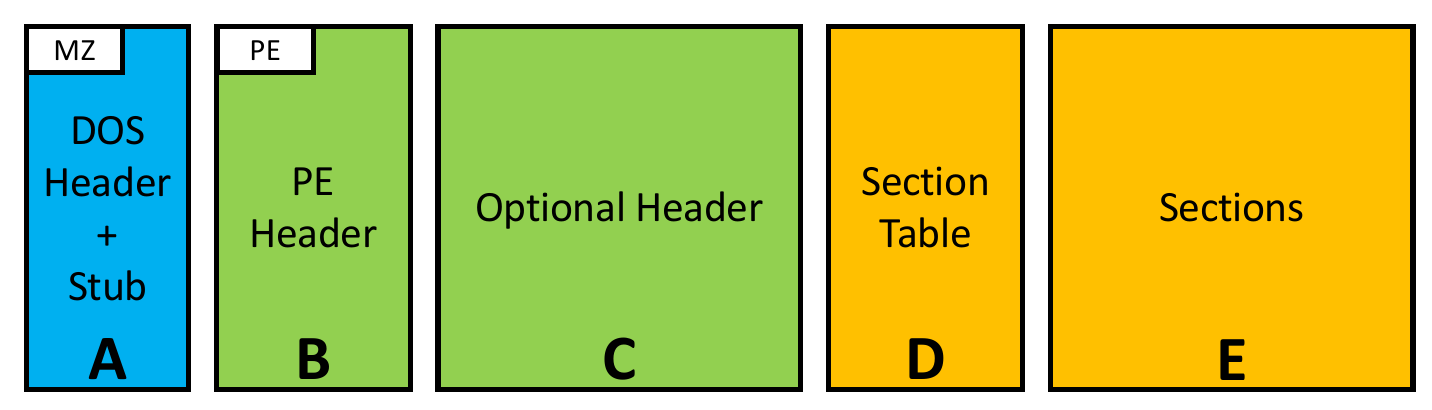}
    \caption[]{The Windows PE file format.
    Each colored section highlights a specific component.
    }
    \label{fig:pe_format}
\end{figure}

\mypar{Windows PE File Format.}
It is the proprietary format developed by Microsoft,\footnote{\url{https://learn.microsoft.com/windows/win32/debug/pe-format}} and it describes how Windows programs are stored as files as shown in \autoref{fig:pe_format}.

\mysubpar{DOS Header and Stub (A).} It contains both the metadata for loading the executable inside a DOS environment, and the \emph{DOS stub}, that prints ``\emph{This program cannot be run in DOS mode}'' when executed in DOS. 
These two components have been kept to maintain compatibility with older Microsoft's operating systems. 

\mysubpar{PE Header (B).} It contains the magic number \verb|PE| along with other file characteristics, such as the target architecture, the header size, creation date, and the file attributes.

\mysubpar{Optional Header (C).} It contains the information used by the operating system to load the program.
It contains offsets that point to useful structures, like the Import Address Table, needed to resolve dependencies, and the Export Table offset, needed for offering functionalities to other software.

\mysubpar{Section Table (D).} It is a list of entries that indicates the characteristics of each core component of the program, and where the OS loader should find them inside the file.

\mysubpar{Sections (E).} These contiguous chunks of bytes contain both data and logic of the executable organized as \emph{.text}, the code of the program, \emph{.data}, global variables, \emph{.rdata} which are read-only constants, and others.

\mypar{Static Windows Malware Detection.}
This task focuses on discriminating between leigitimate programs, the goodware, from the malicious and harmful ones, the malware.
Such can be achieved by either (i) static analysis, which gathers information of the analysed samples without executing them; or through (ii) dynamic analysis, which gathers information from the execution of samples.
In this paper, we will refer only to static analysis, by using machine learning models as decision-making algorithms, achieved in two different ways according to the existing literature.
The first technique directly feeds raw binaries to deep neural networks, without considering any feature extraction phase, achieved by collecting vast corpora of data and training a convolutional neural network on top of them~\citep{raff2017malware, krvcal2018deep, coull2019activation}. 
In this work, we will leverage MalConv~\citep{raff2017malware}, a deep neural network trained on a vast corpus of both benign and malicious programs, that reads the first 2 MB of each input to produce a classification between legitimate or not.
On the contrary, the second trend is based on extracting features from data, thus feeding models with a more compact intermediate representation.
In this work, we will consider the EMBER feature extraction algorithm~\citep{anderson2018EMBER}, which describes Windows PE binaries with 2381 features.
These are defined as: (i) \emph{general file information}, that scrape general information like the filesize, the number of imported and exported features, the presence of the debug section, resources or other components; (ii) \emph{header information}, that scrape relevant characteristics from both the PE and the Optional header, like the architecture needed by the program to safely run, the size of headers, and other features; (iii) \emph{imported functions}, which are the functions required from the program to run properly, and they are stored as string (library name concatenated to the function name) inside an histogram and compressed with hashing; (iv) \emph{exported functions}, which are the functions that are offerend to other programs, stored in the same ways as the imported ones; (v) \emph{section information} that stores the properties of each section like their size, their names, and other features; (vi) \emph{byte histogram} which stores the count of each one of the 256 bytes; (vii) \emph{byte-entropy histogram} that stores the entropy of contiguous chunks of bytes, calculated through sliding windows; (viii) \emph{string information} which stores all the printable strings that are longer than five characters, compressed through hashing.
All this information is used to train a Gradient Boosting Decision Tree (GBDT)~\citep{ke2017lightgbm}.

\mypar{Regression of Malware Detectors.} 
Usually, developers update machine learning models to increase their predictive capabilities at test time, by either applying fine-tuning on newly-available data or re-training models from scratch.
However, even if the overall accuracy increases, some samples that were correctly classified in the past can be mislabelled by the updated model.
We refer to this phenomenon as \emph{regression}~\citep{yan2021positive}, and it is crucial in the domain of malware detection.
In particular, well-known malware samples can bypass the detection\footnote{\url{https://www.mandiant.com/resources/blog/churning-out-machine-learning-models-handling-changes-in-model-predictions}} due to the frequent retraining of machine learning models.
As a consequence, clients can be exposed to older threats that are already available to attackers in the wild.
We refer to this type of regression as \emph{malware flip} (\malflip) throughout the paper.
Lastly, regression can also increase the number of false alarms, since uncontrolled updates might cause benign software to be now detected as malicious.
This regression is not only annoying at a user-experience level, but it could also cause catastrophic failures of the operating system as the updated model is flagging critical software like system utilities as unwanted.
We refer to this type of regression as \emph{goodware flip} (\goodflip) throughout the paper.

\subsection{Adversarial EXEmples} \label{sec:adversarial_EXEmples}
In recent years, we have witnessed the rise of the vulnerabilities of machine learning to \emph{adversarial examples}, carefully crafted samples that induce errors at test time~\citep{biggio2013evasion, biggio2018wild}.
While this was believed to be a problem confined to the domain of computer vision, recent studies showed that it is also possible to create slightly modified Windows malware to evade the detection of machine learning models.
These are the so-called \emph{adversarial EXEmples}~\citep{demetrio2021adversarial}. 
In this work, we selected seven generators of adversarial EXEmples, focusing on the most relevant techniques that exploit all the proposed manipulations~\citep{demetrio2021functionality}.
The generators share a common workflow that consists of using unmodified malware that is perturbed until achieving either evasion or satisfying a stopping condition.
All the manipulations (also referred to as adversarial perturbations) are \emph{practical}, implying that their application is neither corrupting the input sample nor modifying its functionalities~\citep{demetrio2021functionality, demetrio2021adversarial, kozak2023creating}.

\mypar{Partial-, Full-, Extend-DOS Manipulators.}
Demetrio et al.~\citep{demetrio2019explaining, demetrio2021functionality} present different generators of adversarial EXEmples by modifying both the different headers of programs:
(i) the \emph{Partial-DOS} generator that alters the content of the MS-DOS header by manipulating the first 58 bytes after the \verb|MZ| magic number~\citep{demetrio2019explaining};
(ii) the \emph{Full-DOS} generator that re-writes all the DOS header and stub~\citep{demetrio2021adversarial}, thus editing 250 bytes on average; 
and (iii) the \emph{Extend-DOS} generator that enlarges the size of the DOS header and stub by a personalized amount, and then it proceeds by editing all those bytes as the Full-DOS generator~\citep{demetrio2021adversarial}.
These generators have been evaluated with a \emph{gradient-based} optimization algorithm to create adversarial EXEmples against MalConv~\citep{demetrio2021adversarial, demetrio2019explaining}.
In their original evaluations, both Partial- and Full-DOS generators achieve roughly 80\% of evasions, while the Extend-DOS generator outperforms the other generators by bypassing MalConv 93\% of the time.
In this work, we use the implementations of these generators provided inside the SecML Malware library~\citep{demetrio2021secmlmalware}.

\mypar{GAMMA: Injection of Benign Content.}
Demetrio et al.~\citep{demetrio2021functionality} present the Genetic Adversarial Machine Learning Malware Attack (GAMMA), which is a generator that injects content extracted from goodware as new sections of input malware samples.
The attack is defined as a constrained optimization problem solved through the application of a genetic algorithm that maximizes the evasion rate of adversarial EXEmples while minimizing the size of the injected perturbation.
The authors evaluated GAMMA against both GBDT and MalConv models trained on the EMBER dataset, bypassing those detectors on average 69\% and 96\% of cases, respectively.
In this work, we use the implementation of GAMMA provided by the SecML Malware library~\citep{demetrio2021secmlmalware}.

\mypar{Fast Gradient Sign Method (FGSM)}
Kreuk et al.~\citep{kreuk2018deceiving} propose a generator that simultaneously edits content in unused space between sections, namely the \emph{slack} space, or appends content at the end of the file, namely \emph{padding}.
The generator uses the Fast Gradient Sign Method (FGSM)~\citep{goodfellow2015explaining} to optimize the included adversarial content.
The authors test this attack against MalConv, showing that it can be evaded over 99\% of the time.
In this work, we use the implementation of FGSM provided by the SecML Malware library~\citep{demetrio2021secmlmalware}.

\mypar{Gym-Malware}
Anderson et al.~\citep{anderson2018learning} propose a generator that creates adversarial EXEmples by perturbing fields in the Optional Header, adding sections with random content, appending random sequences of bytes, or \emph{packing} the input, which compresses the original program.
This generator leverages a RL algorithm to optimize the sequence of perturbation to apply to malware to make it invisible to the target detector.
Gym-Malware is originally trained and tested against a GBDT trained on a proprietary dataset, achieving an evasion rate between 10\% and 24\%.
The Gym-Malware source codes are available on GitHub repository.\footnote{\url{https://github.com/endgameinc/gym-malware}}

\mypar{MAB-Malware}
Song et al.~\citep{song2022mab} propose a generator that modifies programs by using both the manipulations of Anderson et al.~\citep{anderson2018learning} referred to as \emph{macro manipulations} and by \emph{micro manipulations} which consist of single-byte content editing or injection. 
Similarly to Anderson et al.~\citep{anderson2018learning}, MAB-Malware leverages a RL algorithm to learn the best combination of manipulations to achieve evasion.
Once adversarial EXEmples are found, the generator tries to eliminate unnecessary modification while reducing the size of the manipulation, keeping its stealthiness to the target detector.
Such a minimization process is possible because the application of adversarial perturbation does not consider the sequence of actions, hence some of them can be removed a posteriori.
MAB-Malware is evaluated against both GBDT and MalConv trained on the EMBER dataset and also against commercially available antivirus engines hosted on VirusTotal.
MAB-Malware is shown to successfully bypass the detection of MalConv and GBDT classifiers in over 97\% and 74\% of cases, respectively, and also real-world AVs in up to 48\% of cases. 
In this work, we use the implementation provided by the authors of MAB-Malware,~\footnote{\url{https://github.com/weisong-ucr/MAB-malware}} staging attacks against both MalConv and GBDT.

\mypar{Adversarial Malware Generator (AMG).}
Kozak et al. \citep{kozak2023creating} propose a similar generator to those by Song et al.~\citep{song2022mab} and Anderson et al.~\citep{anderson2018learning}, leveraging RL algorithm to incrementally modify malware samples to create adversarial EXEmples. 
Unlikely Gym-Malware, the primary focus of AMG is the creation of valid adversarial EXEmples that preserve the original functionality. 
Based on this condition, the modifications of Anderson et al.~\citep{anderson2018learning} were reimplemented and extended to better comply with the structure of the PE file format.
To verify the validity of generated EXEmples, the authors designed an algorithm for evaluating binary manipulations inside a sandbox environment, highlighting that the reimplemented AMG perturbations better preserve the original functionality than the previous works.
The authors of AMG used a proximal policy optimization (PPO) agent, trained against the GBDT classifier, that achieves a 53\% evasion rate.
Kozak et al. showed that the performance of the PPO agent is significantly worse when evaluated against commercially-available AVs and that random application of AMG modifications generates more evasive EXEmples than trained counterpart.
In this paper, we use the implementations provided by the authors of AMG.\footnote{\url{https://github.com/matouskozak/AMG}}

\mypar{Defenses against EXEmples}
While plenty of techniques have been proposed to attack Windows malware detectors, the research on defenses against Adversarial EXEmples is still quite unexplored.
Currently, in this domain, the main defensive technique is \emph{adversarial training (AT)}~\citep{madry2018towards, lucas2023adversarial}, which hardens machine learning models by including adversarial EXEmples created with one or more generators inside the training data.
In this work we will consider two different settings for this technique: (i) \emph{incremental} that is applied as a fine-tuning of a baseline model; and (ii) \emph{from scratch}, where a new baseline is trained on all the regular programs combined with adversarial EXEmples.

\section{\exescanner: a Robust Plugin for AVs} 
\begin{figure}[h!]
    \centering
    \includegraphics[width=\linewidth]{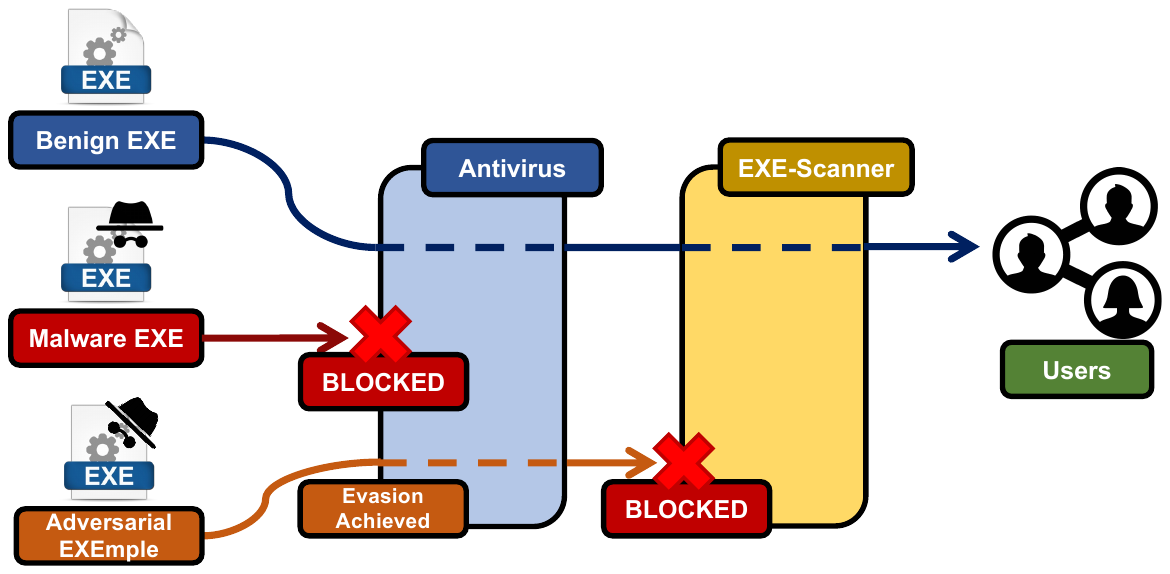}
    \caption{\exescanner in action: regular malware are blocked by the baseline AV, while adversarial EXEmples able to bypass the baseline are blocked by our detector. Goodware is passed through to avoid raising false alarms.}
    \label{fig:EXE-scanner}
\end{figure}

We propose \exescanner, an adversarial EXEmples detection technique designed to serve as an additional line of defense to protect AVs against adversarial EXEmples by rejecting suspicious samples, as depicted in \autoref{fig:EXE-scanner}. 
\exescanner processes the stream of data coming from the defended model, and the samples predicted as benign are double-checked to assess that they are effectively goodware and not adversarial EXEmples.

Since the \exescanner detector should learn to distinguish between benign and potentially adversarial EXEmples, we train the \exescanner on a set of goodware and adversarial EXEmples. 
Due to the proposed design in which the \exescanner is verifying only benign predictions, we omitted training the \exescanner using malware files without adversarial perturbations as these samples should be correctly classified as malicious by the main classifier. 
This decision allows us to create a smaller model that can be easily trained and deployed alongside many anti-malware systems. 
Based on local measurements, a gradient boosting decision tree model described in \autoref{sec:proposed_method_exe-scanner_architectures}, trained on an additional 26531 unmodified malware samples used to generate adversarial EXEmples from train set in \autoref{table:dataset_distribution}, has approximately 32\% larger size on disk and 9.5\% slower training time than the same model trained only on EXEmples and goodware.
Additionally, not including unmodified malware allows us to put more emphasis on the selected generators of adversarial EXEmples, as we are not polluting the training set with unmodified malware that is presumably already detected by the main model.

\subsection{Dataset Creation}
\label{sec:proposed_method_dataset}
To train \exescanner, we need to create a dataset containing adversarial EXEmples computed with most of the generators from the state of the art and extract relevant features from it.
As feature representation, in this work, we use the algorithm presented by Anderson et al.~ \citep{anderson2018EMBER} to transform all binary files into vectors of 2381 features (as described in \autoref{sec:malware_detection}).
In total, we work with three distinct classes of samples.
Firstly, we downloaded three sets of malicious binaries (identified by suffixes 00450, 00454, and 00476) from VirusShare\footnote{\url{https://virusshare.com}} and another set (identified by the name InTheWild.0138) Vx-Underground\footnote{\url{https://vx-underground.org}} online repositories. 
The downloaded malware files were processed using the \textit{pefile}~\citep{Carrera_Ventura_pefile_2023} Python library, and only files that were handled correctly were used in future steps. 
This step was taken to prevent potential future issues with generating adversarial EXEmples. 

\begin{figure}[h!]
    \centering
    \includegraphics[width=0.8\linewidth]{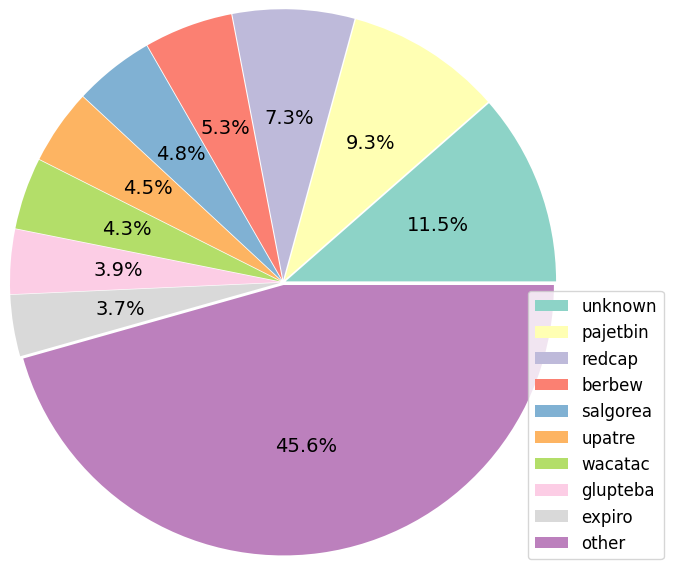}
    \caption{Distribution of malware families inside the set of unmodified malicious files. 
    We flag as \emph{Unknown} all those samples for which AVClass was unable to provide an answer, while we group as \emph{other} the remaining family with less than 2.0\% of presence in our dataset.}
    \label{fig:dataset_composition}
\end{figure}

We selected 30000 malicious binaries from the VirusShare samples (10000 from each set) and 1000 malicious samples from the Vx-Underground set.
We label these samples as \textit{unmodified} malware as we consider this set not to contain perturbations introduced by generators of adversarial EXEmples. 
Using VirusTotal\footnote{\url{https://www.virustotal.com}} API and AVClass\footnote{\url{https://github.com/malicialab/avclass}} tool, we collected a distribution of malware families shown in \autoref{fig:dataset_composition}.

The set of unmodified malware serves as the foundation for creating the second class representing adversarial EXEmples.
EXEmples are created by executing the generators presented in \autoref{sec:adversarial_EXEmples}, configured as follows:
\begin{itemize}
    \item \textbf{PartialDOS.} We run the attack against the MalConv classifier contained in the library, using the default configuration coded in SecML Malware (which perturbs 58 bytes), optimizing the sample for 10 iterations;
    \item \textbf{FullDOS.} We run the attack against the MalConv classifier contained in the library, using the default configuration coded in SecML Malware (which perturbs all the header, counting on average 250 bytes), optimizing the sample for 10 iterations;
    \item \textbf{ExtendDOS.} We run the attack against the MalConv classifier contained in the library, using the default configuration coded in SecML Malware (which extends the DOS header by 512 bytes, adjusted to match the file alignment of the sample), optimizing the sample for 10 iterations;
    \item \textbf{GAMMA.} We rely on the section-injection attack by injecting 10 sections randomly extracted from a pool of 1500 goodware files. We set the number of queries to 100, the penalty term on the size is set to $10^{-6}$, and we target the MalConv classifier contained in the library;
    \item \textbf{FGSM.} We run the attack to inject 2048 padding bytes and fill the slack space between sections, iterating five times and targeting the MalConv classifier contained in the library;
    \item \textbf{Gym-Malware.} We use the default configuration found in the official implementation developed by the authors, trained using 4000 malicious samples over 50000 iterations. We target the GBDT classifier, trained on 100000 samples, provided by the authors. We consider two settings: SC-Gym-Malware that uses the scores of a model for optimizing manipulations, and BB-Gym-Malware that attacks a target classifier in a black-box fashion;
    \item \textbf{MAB-Malware.} We setup the provided Docker container to run the attack against both MalConv (MAB-MalConv) and GBDT trained on EMBER (MAB-EMBER) using the default configuration provided by the authors of the technique;
    \item \textbf{AMG.} We use the pre-trained PPO policy (AMG-PPO) from the authors~\citep{kozak2023creating} and the policy that performs random actions (AMG-Random).
    Both are executed to bypass the GBDT trained on the EMBER dataset.
\end{itemize}

\noindent Note that not all malware files were possible to adversarially perturb by the selected generators. Consequently, we discarded samples that did not contain any modifications.

The last class, \textit{goodware} files, is formed by a set of 1592 executables that were taken from fresh Windows 10 installation, a set of 5000 goodware binaries that were scrapped from GitHub public repositories, and a set of over 47000 goodware programs scrapped from Chocolatey\footnote{\url{https://chocolatey.org}} package manager.
On top of these samples, we use 100000 goodware feature vectors from the EMBER test set.
All the data we use are publicly available either as a part of the EMBER dataset,\footnote{\url{https://github.com/elastic/ember}} or from our repository.\footnote{\url{https://github.com/matouskozak/EXE-scanner}}
We verified all unmodified malware samples with 10 commercially available AVs, such as Avast, Microsoft Defender, ESET, or Kaspersky, to ensure they were truly malicious samples, and we discarded the ones that were labeled by more than five AVs as harmless. 
The AV programs were selected by considering leading anti-malware products based on the annual 2022 Summary Report of AV-Comparatives~\citep{avcomparatives2022summary}. 
We also eliminated corresponding EXEmples that were created from discarded unmodified malware, and similarly, we removed benign samples that were considered harmful by more than five AVs. 

\mypar{Time analysis.}
Even though every PE file contains a field inside the PE header named \verb|TimeDateStamp| (\autoref{sec:background}) containing a UNIX timestamp of its creation, such a piece of information can be arbitrarily changed by the author.
Consequently, to have a reliable source of time origin information for the collected malware and goodware samples, we use VirusTotal's ``First Submission'' attribute. 
The oldest file in our dataset date back to May of 2006, with the newest coming from December of 2024. 
Based on the gathered time information, we decided to split our dataset into two main subsets: (i) samples from the years 2006-2021 and (ii) samples from the years 2022-2024. 
The subset (i) was further split into train and test sets.
This time separation allows us to study how models perform on current data originating from the same time distribution as the training set and on future data when models are not updated.
The overall distribution of samples between train and $\text{test}_1$ (i), and $\text{test}_2$ (ii) is shown in \autoref{table:dataset_distribution}.

\begin{table}[h!]
    \centering
    \caption{Distribution of collected goodware, unmodified malware, and adversarial EXEmples samples between train, $\text{test}_1$, and $\text{test}_2$ sets.}
    \label{table:dataset_distribution}
    \resizebox{\columnwidth}{!}{%
    \begin{tabular}{@{}lcccc@{}}
        \toprule
        & \textbf{Goodware} & \textbf{Unmodified Malware} & \textbf{Adv. EXEmples} & \textbf{Total}  \\ \midrule
        Train ($< 2022$)     & 100000 & -        & 119411   & \textbf{219411} \\
        $\text{Test}_1$ ($< 2022$)     & 24202   & 1930        & 14393    & \textbf{40525}  \\ 
        $\text{Test}_2$ ($\geq 2022$)     & 30514   & 1007         & 8076    & \textbf{39597}  \\ \bottomrule
    \end{tabular}
    }
\end{table}

\subsection{\exescanner Architecture} \label{sec:proposed_method_exe-scanner_architectures}
We envision that different classifier architectures could be used to implement the concept of \exescanner presented in \autoref{fig:EXE-scanner}. In this work, we consider three classifiers: (i) logistic regression; (ii)  support vector machine; and (iii) gradient boosting decision tree.

We use logistic regression for binary classification trained for up to 10000 iterations using Scikit-learn \citep{scikit-learn} library with the default configuration. 
For faster convergence, we standardize features by removing the mean and dividing with the standard deviation computed for each feature.

Next, we use a support vector machine in the form of a support vector classifier (SVC) provided by Scikit-learn \citep{scikit-learn} library with the default configuration.
Same as with logistic regression, we standardize features by removing the mean and divding with standard deviation for faster convergence.

The gradient boosting decision tree (GBDT) used in this work is based on the architecture of the GBDT classifier presented by \cite{anderson2018EMBER}. 
This model is a gradient boosting decision tree with 2048 leaves, a maximum depth of 15 for individual trees, trained over 1000 iterations using the LightGBM \citep{ke2017lightgbm} framework with a 0.05 learning rate.

\section{Experimental Analysis} 
\label{sec:new_experiments}
We evaluate \exescanner performance with standalone classifiers, compare it to adversarial training, and test its robustness against new adversarial EXEmples. 

In particular, we will analyze the following aspects: 
(i) how different \exescanner models perform (\autoref{sec:evaluation_exe_scanner_architectures}); 
(ii) how much \exescanner can harden a model trained on a large corpus of data against \emph{transfer attacks} (\autoref{sec:transfer}); 
(iii) how well \exescanner is able to mitigate the effectiveness of EXEmples proposed against the original target classifiers (\autoref{sec:hardening_models_with_exescanner}); 
(iv) what features \exescanner utilizes to detect adversarial EXEmples (\autoref{sec:evaluation_explainability}); 
(v) how \exescanner performs when compared to \emph{adversarial training} (\autoref{sec:at_exescanner}); 
(vi) how \exescanner can be used as a plugin for commercially-available AVs (\autoref{sec:av_exescanner}); 
(vii) how \exescanner can withstand adversarial EXEmples directly computed against it in an adaptive fashion (\autoref{sec:exescanner_robustness}); 
and (viii) how \exescanner and AT performance deteriorates over time when not updated.

\subsection{Experimental Setup}
\label{sec:evaluation_setup}

\mypar{Metrics.} For the evaluation of \exescanner, we use well-known metrics such as accuracy, precision, recall (true positive rate, TPR, detection rate), balanced F-score (F1), and false positive rate (FPR).
In particular, we will refer to ``recall'' as the recall computed on all data (both unmodified malware and adversarial EXEmples), to ``recall (adv.)'' as the recall computed \emph{only} on adversarial EXEmples, and to ``recall (unm.)'' as the recall computed \emph{only} on unmodified malware. 
When considering updating existing malware detectors, we also measure their accuracy regression~\citep{yan2021positive} as described in \autoref{sec:malware_detection}, reporting both the malware (\malflip) and the goodware flips (\goodflip). If not stated otherwise, the following results are rounded to two decimal places.

\mypar{Hardware.} Experiments and generation of datasets were performed on the NVIDIA DGX Station A100 workstation fitted with four NVIDIA A100 40GB graphic cards, a single AMD Epyc 7742 processor, and 512 GB of system memory. 
However, all experiments are also reproducible on a single-processor workstation with a sufficient amount of system memory (32 GB recommended).

\mypar{Baseline model.} For evaluating the efficacy of \exescanner or AT, we train a baseline model to harden, which is the GBDT model trained on the EMBER train dataset \citep{anderson2018EMBER} (samples from years 2006 to 2018). 
We refer to this model as ``Baseline''.

\subsection{\exescanner Architectures}
\label{sec:evaluation_exe_scanner_architectures}
\begin{figure}[h!]
    \centering
    \includegraphics[width=0.5\textwidth]{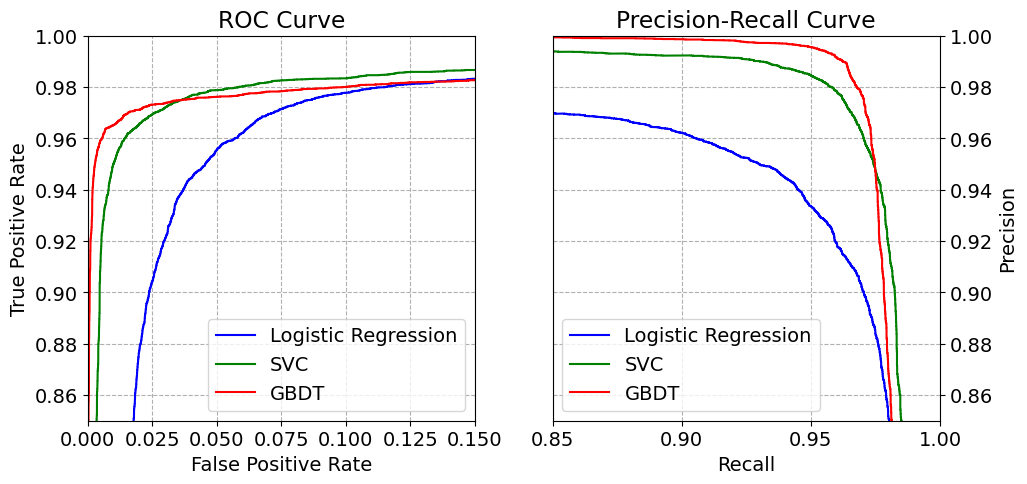}
    \caption{ROC (on the left) and precision-recall (on the right) curves for logistic regression, SVC, and GBDT \exescanner{}s on the $\text{test}_1$ set.}
    \label{fig:exescanner_architecture_comparison}
\end{figure}

At first, we compare the potential of different classifiers used as \exescanner plugin. 
We consider logistic regression, SVC, and GBDT classifiers described in \autoref{sec:proposed_method_exe-scanner_architectures} trained on the training dataset from \autoref{table:dataset_distribution}.
We evaluate the performance of all three classifiers in isolated settings without the main classifier on the $\text{test}_1$ set. 
In \autoref{fig:exescanner_architecture_comparison}, we display the ROC and precision-recall curves for the classifiers.

Note this comparison is not a real-world scenario since \exescanner is intended to be used as a plugin classifier. 
However, based on the curve trends, we can determine which classifier has a bigger potential in blocking most adversarial EXEmples. 
For the malware detection domain, where models operate at very low levels of FPR (1\% or lower), we can see that the GBDT model is more suitable as it achieves higher TPR at low FPR as well as maintaining both high precision and recall. 
In the following experiments, we consider GBDT as the default architecture for \exescanner.

\subsection{Transfer Attacks against \exescanner}
\label{sec:transfer}
Now, we want to assess the performance of \exescanner in an environment when a classifier, trained from scratch on a large dataset, behaves against \emph{transfer attacks}~\citep{demontis2019adversarial}.
These are adversarial EXEmples optimized against a specific malware detector and then tested against another one.
Intuitively, transfer attacks might be effective against detectors built using similar models and trained on similar datasets.
Also, these attacks are the easiest to stage against unknown models since an attacker can optimize adversarial EXEmples against a classifier they own at no cost~\citep{demetrio2021functionality}.
Hence, we mimic this setting by using the aforementioned baseline GBDT classifier paired with trained \exescanner.
We configure the threshold for the baseline GBDT and the \exescanner to score 1\%, 0.1\%, and 0.01\% FPR on the $\text{test}_1$ set. 
\exescanner's threshold is set after the baseline model's threshold is determined in a way that all malicious predictions are fixed, and benign predictions are replaced with probabilities reported by \exescanner.
As a result, we simulate a scenario where the main classifier is independent of \exescanner, which is distributed as a plugin defense mechanism.
We proceed by testing baseline's and \exescanner's robustness against adversarial EXEmples, and we report the results in \autoref{table:baseline_EXE-scanner}.

\begin{table*}[h!]
    \centering
    \caption{Performance of our baseline GBDT classifier expanded with \exescanner at 1\% (\ref{subtable:baseline_EXE-scanner_1}), 0.1\% (\ref{subtable:baseline_EXE-scanner_0.1}), and 0.01\% (\ref{subtable:baseline_EXE-scanner_0.01}) FPR levels on the $\text{test}_1$ set.}
    \label{table:baseline_EXE-scanner}
    \begin{subtable}{0.33\textwidth}
        \centering
        \resizebox{0.9\linewidth}{!}{\begin{tabular}{@{}lcc@{}}
            \toprule
                                & \textbf{Baseline} & \textbf{\exescanner} \\ \midrule
            Accuracy             & 95.83    & 98.09       \\
            Precision            & 98.4     & 98.49       \\
            Recall               & 91.14    & 96.73       \\
            Recall (unm.)        & 90.88    & 92.69       \\
            Recall (adv.)        & 91.18    & 97.27       \\
            F1                   & 94.63    & 97.60       \\ \bottomrule
        \end{tabular}}
    \caption{1\% FPR}
    \label{subtable:baseline_EXE-scanner_1}
    \end{subtable}%
    \begin{subtable}{0.33\textwidth}
        \centering
        \resizebox{0.9\linewidth}{!}{\begin{tabular}{@{}lcc@{}}
            \toprule
                            & \textbf{Baseline} & \textbf{\exescanner} \\ \midrule
            Accuracy        & 80.75    & 90.23       \\
            Precision       & 99.72    & 99.81       \\
            Recall          & 52.36    & 75.90       \\
            Recall (unm.)   & 52.44    & 59.95       \\
            Recall (adv.)   & 52.35    & 78.04       \\
            F1              & 68.67    & 86.23       \\ \bottomrule
        \end{tabular}}
    \caption{0.1\% FPR}
    \label{subtable:baseline_EXE-scanner_0.1}
    \end{subtable}%
    \begin{subtable}{0.33\textwidth}
        \centering
        \resizebox{0.9\linewidth}{!}{\begin{tabular}{@{}lcc@{}}
            \toprule
                            & \textbf{Baseline} & \textbf{\exescanner} \\ \midrule
            Accuracy        & 72.84    & 87.21       \\
            Precision       & 99.96    & 99.98       \\
            Recall          & 32.59    & 68.26       \\
            Recall (unm.)   & 37.46    & 47.05       \\
            Recall (adv.)     & 31.94    & 71.10       \\
            F1              & 49.16    & 81.13       \\ \bottomrule
        \end{tabular}}
    \caption{0.01\% FPR}
    \label{subtable:baseline_EXE-scanner_0.01}
    \end{subtable}%
\end{table*}

The ``baseline'' and ``\exescanner'' columns represent the performance of the baseline GBDT classifier and baseline GBDT classifier extended by \exescanner, respectively.
As expected, transfer attacks are not very effective against a never-seen model (over 91\% recall at 1\% FPR for baseline classifier displayed in \autoref{subtable:baseline_EXE-scanner_1}), as already shown in previous work~\citep{demetrio2021adversarial}.
However, even in this case, some adversarial EXEmples are able to bypass the detection of the baseline model, which are promptly stopped by \exescanner, improving the detection rate of adversarial EXEmples of the combined system to over 97\%.
For lower levels of FPR (0.1\% and 0.01\% shown in \autoref{subtable:baseline_EXE-scanner_0.1} and \autoref{subtable:baseline_EXE-scanner_0.01}, respectively), we can observe a rapid decrease in recall performance of the baseline model.
While baseline model recall decreases for both unmodified and adversarial EXEmples of malware, \exescanner extended detector records significantly smaller decrease, especially in the detection rate of adversarial EXEmples.

Further, we examine which adversarial EXEmples are able to sneak under the radar by presenting a detailed overview of EXEmples' evasion rates in \autoref{table:evasion_rates_generators}. 
The table depicts evasion rates of adversarial EXEmples from different generators against the baseline and \exescanner enhanced malware classifiers at 1\% FPR. 

\begin{table}[h!]
    \centering
    \caption{Evasion rates of adversarial EXEmples from different generators between baseline and \exescanner enhanced GBDT malware classifiers on the $\text{test}_1$ set.}
    \label{table:evasion_rates_generators}
    \begin{tabular}{@{}lccc@{}}
        \toprule
        ~                     & \textbf{Baseline} & \textbf{EXE-scanner} & \textbf{Decrease}\\ \midrule
        PartialDOS            & 10.6     & 3.8         & 64.15                \\
        FullDOS               & 8.4      & 3.2         & 61.9                 \\
        ExtendDOS             & 7.96     & 5.24        & 34.23                \\
        FGSM                  & 0.86     & 0.78        & 9.09                 \\
        AMG-PPO               & 12.25    & 0.84        & 93.17                \\
        AMG-Random            & 6.99     & 1.02        & 85.47                \\
        GAMMA                 & 7.61     & 2.66        & 65.09                \\
        SC-Gym-Malware        & 1.63     & 1.09        & 33.33                \\
        BB-Gym-Malwre         & 1.43     & 1.09        & 23.81                \\
        MAB-MalConv           & 7.35     & 3.6         & 51.0                 \\
        MAB-EMBER             & 28.59    & 7.5         & 73.75                \\ \bottomrule
    \end{tabular}
\end{table}

Based on the presented results, we can conclude that the presence of \exescanner as a plugin significantly decreases the evasion capabilities of generated adversarial EXEmples.
Evasion rates of all EXEmple generators decreased down to single-digit values or less, with some generators' evasion rates decreasing by more than 70\%.

\subsection{Hardening Models with \exescanner}
\label{sec:hardening_models_with_exescanner}

As a next experiment, we evaluate the efficacy of \exescanner in detecting adversarial EXEmples that were evading the model they have been optimized against, as specified in \autoref{sec:proposed_method_dataset}. 
We use the same \exescanner as in the previous experiment (\autoref{sec:transfer}) with the threshold set on 1\% FPR.
The \exescanner is appended to the target classifier, and we measure the performance on the $\text{test}_1$ set.
The resuls are shown in \autoref{table:patching_classifiers}

The ``Target Classifier'' and ``\exescanner'' columns represent original and \exescanner hardened detection rates of adversarial EXEmples designed to evade the original target classifier.
These EXEmples were computed from the $\text{test}_1$ set of unmodified malware samples, not included inside the training of \exescanner.
Hence, from these results, we can state that \exescanner is able to generalize across EXEmples from different generators, significantly increasing the detection of all target classifiers.
In particular, even in cases where EXEmples from specific generators are detected in less than 10\%, like MAB-MalConv against MalConv, \exescanner is able to mitigate this issue by inflating recall of adversarial EXEmples to $\sim$ 90\%.
This increase is represented in the last column of \autoref{table:patching_classifiers}, where we note the percentage increase of recall thanks to \exescanner.
Lastly, for each attack, we report the rounded average size of the modified and/or injected content by computing the Levenshtein edit distance between the original unmodified malware and the adversarial EXEmple computed from it.
From the first column of \autoref{table:patching_classifiers}, we remark that there are some cases in this domain where higher perturbation size does not lead to a higher evasion rate of adversarial EXEmples.
Rather, we observe that adversarial EXEmples against a specific model might be more effective than others since they exploit known vulnerabilities.
For instance, MalConv is known to be susceptible to attacks that modify the header~\citep{demetrio2019explaining, demetrio2021functionality}, and the perturbation of roughly 120 bytes of FullDOS is more effective than the manipulation of 2048 bytes achieved by FGSM.

\begin{table}[t]
    \centering
    \caption{Detection rates (recall) of adversarial EXEmples of the original target classifier (\textbf{G} stands for GBDT, and \textbf{M} stands for MalConv), and when extended with \exescanner. We also report the average perturbation size computed through edit distance between unmodified malware and their adversarial EXEmple counterparts.}
    \label{table:patching_classifiers}
    \resizebox{\linewidth}{!}{%
    \begin{tabular}{@{}lrcccc@{}}
        \toprule
        ~ & \textbf{Edit Distance} & \textbf{Target Classifier} & \textbf{EXE-scanner} & \textbf{Increase} \\ \midrule
        PartialDOS            & 58      & (M) 67.40             & 87.40       & 20.00     \\
        FullDOS               & 117     & (M) 45.00             & 83.00       & 38.00     \\
        ExtendDOS             & 713     & (M) 57.17             & 90.82       & 33.64     \\
        FGSM                  & 2048    & (M) 58.61             & 87.48       & 28.87     \\
        AMG-PPO               & 129158  & (G) 45.40             & 93.97       & 48.57     \\
        AMG-Random            & 159838  & (G) 62.37             & 93.43       & 31.06     \\
        GAMMA                 & 321978  & (M) 63.96             & 95.69       & 31.73     \\
        SC-Gym-Malware        & 438072  & (G) 93.54             & 94.69       & 1.16      \\
        BB-Gym-Malware        & 433600  & (G) 93.27             & 94.97       & 1.70      \\
        MAB-MalConv           & 898176  & (M) 9.33              & 89.93       & 80.60     \\
        MAB-EMBER             & 2656435 & (G) 26.27             & 76.47       & 50.21     \\ \bottomrule
    \end{tabular}}
\end{table}

\subsection{Explaining the Behavior of \exescanner}
\label{sec:evaluation_explainability}
To understand the detection capabilities of \exescanner shown in \autoref{sec:transfer} (\autoref{table:evasion_rates_generators}) and \autoref{sec:hardening_models_with_exescanner} (\autoref{table:patching_classifiers}), where some EXEmples from specific generators are harder to detect, we conduct an a posteriori analysis of the underlining GBDT classifier serving as \exescanner using the SHAP library \citep{NIPS2017_7062}.

Specifically, we are interested in feature space artifacts left by distinct generators of adversarial EXEmples.
The following are key features learned by \exescanner to detect adversarial EXEmples:
\begin{itemize}
    \item \textbf{PartialDOS} and \textbf{FullDOS}: We were unable to identify any feature correlated with the perturbations introduced by these generators.
    However, both perturbations are small in size, thus they have a negligible impact on the EMBER feature representation used by \exescanner.
    \item \textbf{ExtendDOS}: \verb|SizeOfHeaders| from Optional Header influenced by extending the DOS header and stub that increase the overall header size.
    \item \textbf{FGSM}: Due to the fact that this generator modifies unused space in programs, we can observe spikes in the relevance attributes on the byte histogram features.
    \item \textbf{AMG-PPO}: Features originating from the number of printable strings influenced by modifications, such as appending benign-looking content from goodware binaries to the file overlay.
    \item \textbf{AMG-Random}: Total number of sections influenced by adding new sections that lead to an increase in the section count.
    \item \textbf{GAMMA}: Total number of sections controlled by the section-injection attack that increases the section count.
    \item \textbf{SC-Gym-Malware} and \textbf{BB-Gym-Malware}: Number of sections with read and execute permission influenced by modifications such as adding new ``.text`` sections.
    \item \textbf{MAB-MalConv} and \textbf{MAB-EMBER}: Overall size of the binary influenced by the considerable average perturbation size as reported in \autoref{table:patching_classifiers}.
\end{itemize}

\noindent Note that we omitted features that were shared across all generators, such as \verb|TimeDateStamp|, size of resource table, or \verb|Characteristics| from PE Header that are not specific to individual generators but rather to the original malware used as an input for creation of the EXEmples.
Based on the analysis presented, where \exescanner did not detect significant feature artifacts for the Partial and Full-DOS generators, we can better understand why \exescanner cannot provide robust protection against EXEmples originating from these generators as shown in \autoref{table:evasion_rates_generators} and \autoref{table:patching_classifiers}.
Additionally, learned features such as the overall size of the binary for the MAB-MalConv and MAB-EMBER generators can easily lead to incorrect predictions if a large goodware executable is tested.

\subsection{Security Regression and Comparison with Adversarial Training}
\label{sec:at_exescanner}
As a direct competitor to our approach, we compare our methodology against \emph{adversarial training} (AT), a commonly used technique to improve the robustness against adversarial threats (described in \autoref{sec:adversarial_EXEmples}).
In fact, we envision a scenario where practitioners replace their detectors with either (i) robust fine-tuned versions of them (AT-incremental), computed by performing an additional training step on the same dataset used by \exescanner; or (ii) models re-trained from scratch (AT-full).
Since these act as replacements for models in production, we are interested not only in comparing their predictive capabilities and robustness but also in the regression they face in terms of both malware (\malflip) and goodware (\goodflip) flips (described in \autoref{sec:malware_detection}).
As a baseline model to harden, we leverage the same GBDT model trained in \autoref{sec:evaluation_setup}.
We augment the baseline model's datasets with goodware and adversarial EXEmples used to train \exescanner (see \autoref{table:dataset_distribution}), and we evaluate incremental and from scratch scenarios.

During the training phase, we recorded compute requirements for each of the scenarios and reported the measured values in \autoref{table:compute_requirements}.
The table presents time in seconds (s) and memory in megabytes (MB) required to train the aforementioned model configuration.
The recorded values align with expectations that AT-full is the most resource-demanding of all tested scenarios as it operates with the biggest dataset.
The \exescanner and AT-incremental, using the same dataset, have similar memory requirements, yet the time needed for training is much smaller for AT-incremental.
This can be explained by the fact that AT-incremental performs only fine-tuning update of the baseline model whereas \exescanner trains a new plugin detector.
\begin{table}[h!]
    \centering
    \caption{Time (in seconds) and memory (in MB) needed for training baseline, \exescanner, AT-incremental, and AT-full models.}
    \label{table:compute_requirements}
    \resizebox{\columnwidth}{!}{%
    \begin{tabular}{@{}lcccc@{}}
        \toprule
        ~ & \textbf{Baseline} & \textbf{\exescanner} & \textbf{AT-incremental} & \textbf{AT-full} \\ \midrule
        Time        & 1144  & 290    & 24    &   1374   \\
        Memory      & 22848 & 20014  & 20736 &   25315  \\ \bottomrule
    \end{tabular}
    }
\end{table}

To compare all these different scenarios fairly, we compute detection thresholds of models to score 1\%, 0.1\%, and 0.01\% FPRs on the $\text{test}_1$ set. The results, including \goodflip and \malflip regression values, are shown in \autoref{table:comparison_AE_detector_and_AT}.

\begin{table*}[h!]
    \centering
    \caption{Comparison of incremental and from scratch AT with \exescanner extended malware classifier at 1\% (\ref{subtable:comparison_AE_detector_and_AT_1}), 0.1\% (\ref{subtable:comparison_AE_detector_and_AT_0.1}), and 0.01\% (\ref{subtable:comparison_AE_detector_and_AT_0.01}) FPR levels on the $\text{test}_1$ set. We report both \malflip and \goodflip regressions occurring while updating models.}
    \label{table:comparison_AE_detector_and_AT}
    \begin{subtable}{0.5\textwidth}
        \centering
        \resizebox{0.95\columnwidth}{!}{%
        \begin{tabular}{@{}lcccc@{}}
            \toprule
            ~ & \textbf{Baseline} & \textbf{\exescanner} & \textbf{AT-incremental} & \textbf{AT-full} \\ \midrule
            Accuracy             & 95.83    & 98.09       & 98.35          & 98.58   \\
            Precision            & 98.40    & 98.49       & 98.51          & 98.53   \\
            Recall               & 91.14    & 96.73       & 97.38          & 97.94   \\
            Recall (unm.)        & 90.88    & 92.69       & 94.40          & 95.70    \\
            Recall (adv.)        & 91.18    & 97.27       & 97.78          & 98.24   \\
            F1                   & 94.63    & 97.60       & 97.94          & 98.23   \\ \midrule
            \goodflip            & -        & 0           & 56             & 154     \\
            \malflip             & -        & 0           & 31             & 169     \\ \bottomrule
        \end{tabular}
        }
    \caption{1\% FPR}
    \label{subtable:comparison_AE_detector_and_AT_1}
    \end{subtable}%
    \begin{subtable}{0.5\textwidth}
        \centering
        \resizebox{0.95\columnwidth}{!}{%
        \begin{tabular}{@{}lcccc@{}}
            \toprule
            ~ & \textbf{Baseline} & \textbf{\exescanner} & \textbf{AT-incremental} & \textbf{AT-full} \\ \midrule
            Accuracy             & 80.75    & 90.23       & 85.88          & 94.48   \\
            Precision            & 99.72    & 99.81       & 99.77          & 99.83   \\
            Recall               & 52.36    & 75.90       & 65.10          & 86.44   \\
            Recall (unm.)        & 52.44    & 59.95       & 68.03          & 82.44   \\
            Recall (adv.)        & 52.35    & 78.04       & 64.71          & 86.98   \\
            F1                   & 68.67    & 86.23       & 78.79          & 92.66   \\ \midrule
            \goodflip            & -        & 0           & 8              & 16      \\
            \malflip             & -        & 0           & 64             & 630     \\ \bottomrule
        \end{tabular}
        }
    \caption{0.1\% FPR}
    \label{subtable:comparison_AE_detector_and_AT_0.1}
    \end{subtable}%
    \\
    \begin{subtable}{0.5\textwidth}
        \centering
        \resizebox{0.95\columnwidth}{!}{%
        \begin{tabular}{@{}lcccc@{}}
            \toprule
            ~ & \textbf{Baseline} & \textbf{\exescanner} & \textbf{AT-incremental} & \textbf{AT-full} \\ \midrule
            Accuracy             & 72.84    & 87.21       & 75.11          & 72.38   \\
            Precision            & 99.96    & 99.98       & 99.97          & 99.96   \\
            Recall               & 32.59    & 68.26       & 38.23          & 31.45   \\
            Recall (unm.)        & 37.46    & 47.05       & 40.26          & 24.97   \\
            Recall (adv.)        & 31.94    & 71.10       & 37.96          & 32.32   \\
            F1                   & 49.16    & 81.13       & 55.31          & 47.85   \\ \midrule
            \goodflip            & -        & 0           & 1              & 2       \\
            \malflip             & -        & 0           & 242            & 2305    \\ \bottomrule
        \end{tabular}
        }
    \caption{0.01\% FPR}
    \label{subtable:comparison_AE_detector_and_AT_0.01}
    \end{subtable}%
\end{table*}

At first glance, we can see a vast difference in overall accuracy and robustness to adversarial EXEmples at different FPR levels.
At 1\% FPR, our results show that the best model w.r.t. robustness to adversarial EXEmples is AT-full, closely followed by AT-incremental and \exescanner.
Intuitively, the fine-tuning of the GBDT model might miss some correlations with the unmodified data that are only provided to AT-full instead. 
Hence, this might result in a drop in performance with respect to \exescanner and AT-incremental, even if they share the same adversarial EXEmples in the training set.
However, we highlight two downsides of considering the update through retraining incrementally and from scratch.
Firstly, even if AT-full exhibits improved predictive capabilities (up to 98.58\% at 1\% FPR), the model is facing a regression in the accuracy of malware samples by counting 169 malware samples at 1\% FPR that are now mispredicted as goodware.
\exescanner, on the other hand, due to its design where only benign predictions by the main classifier are passed to the plugin (see \autoref{fig:EXE-scanner}), cannot introduce \malflip{}s.

Secondly, both AT-incremental and AT-full are unable to correctly classify 56 and 154 goodware that were previously correctly classified.
This might be caused by EXEmples inside the training set that resemble benign programs by either modifying only a few bytes in headers or by injecting content from the goodware distribution~\citep{demetrio2021functionality}.
Hence, some benign programs might be mislabelled as malicious since they are too similar to adversarial EXEmples, potentially provoking serious issues if these models get deployed as they might stop relevant Windows components from running, causing either crashes at the operating-system level or annoying behavior that prevents programs wanted by users from running properly.

At 0.1\% FPR, AT-full is the most robust model, followed by \exescanner and AT-incremental.
As expected, we see for all models drops in recall performance coupled with improved precision at lower FPR.
The biggest drop in performance, over 12\% in accuracy and 32\% in recall, is recorded for AT-incremental.
Similarly to the 1\% FPR scenario, both AT-incremental and AT-full experience goodware and malware regressions. On the other hand, \exescanner extended baseline model maintains its correct prediction behavior while improving robustness against adversarial EXEmples by over 25\%.

At 0.01\% FPR, the performance of both AT-incremental and AT-full deteriorates significantly, offering only marginal improvements (or none in the case of AT-full) over the baseline model.
\exescanner, in contrast, continues to offer protection against the majority of malicious adversarial EXEmples with a recall of 71.10\%.

To match match real-world settings where malware are much more rare than goodware samples~\citep{pendlebury2019tesseract}, we also provide evaluations on test$_1$ by imposing a 10:1 ratio between goodware and malware.
Such is achieved by randomly removing malware from such a test set to match this criterion and keeping all goodware samples, and we report all the results in \autoref{table:comparison_AE_detector_and_AT_10:1} in \autoref{appendix:B}.
Our findings are aligned with those on \autoref{table:comparison_AE_detector_and_AT}, confirming once more the effectiveness of \exescanner.

These results convey that \exescanner can achieve similar robustness and accuracy as adversarial training without incurring regression of performances.
For these two reasons, security engineers might opt to employ \exescanner-like defense method since it better preserves the original prediction behavior while enhancing robustness against adversarial EXEmples, even at low levels of FPR.

\subsection{Pairing \exescanner with Commercially AVs}
\label{sec:av_exescanner}

We now show the efficacy of \exescanner when it is paired with commercially-available AV engines.
We select the same 10 top-tier AV engines we used for described in \autoref{sec:proposed_method_dataset} hosted on the cloud service VirusTotal, couple them with \exescanner used in \autoref{sec:transfer}, and we test the combination with goodware, unmodified malware, and the adversarial EXEmples from $\text{test}_1$ set.
To be compliant with the policy and terms of use of VirusTotal, we anonymize all the results of the considered AV engines, thus avoiding spreading unwanted comparisons between products.

\begin{table*}[h!]
    \centering
    \caption{Performance of \exescanner (0.01\% FPR threshold) with different AV engines on $\text{test}_1$ set.}
    \label{table:distributed_EXE-scanner_AVs_0.01}
    \resizebox{\textwidth}{!}{%
    \begin{tabular}{@{}l|cc|cc|cc|cc|cc@{}}
    \toprule
    ~ & \textbf{AV-1}  & \textbf{\exescanner} & \textbf{AV-2}  & \textbf{\exescanner} & \textbf{AV-3}  & \textbf{\exescanner} & \textbf{AV-4}  & \textbf{\exescanner} & \textbf{AV-5}  & \textbf{\exescanner} \\ \midrule
    Accuracy             & 94.18 & 97.65             & 94.68       & 97.73                   & 92.59     & 96.75                 & 93.95 & 96.63             & 94.96 & 97.81             \\
    Precision            & 99.74 & 99.76             & 99.77       & 99.79                   & 99.63     & 99.67                 & 99.83 & 99.85             & 99.26 & 99.32             \\
    Recall               & 85.77 & 94.39             & 86.99       & 94.58                   & 81.91     & 92.23                 & 85.13 & 91.77             & 88.15 & 95.21             \\
    Recall (unm.)        & 98.91 & 98.96             & 97.88       & 97.98                   & 96.32     & 96.37                 & 94.87 & 94.97             & 98.13 & 98.24             \\
    Recall (adv.)        & 84.01 & 93.77             & 85.53       & 94.12                   & 79.98     & 91.68                 & 83.83 & 91.34             & 86.81 & 94.80              \\
    F1                   & 92.23 & 97.00             & 92.94       & 97.11                   & 89.90     & 95.81                 & 91.90 & 95.64             & 93.37 & 97.22             \\
    FPR                  & 0.15  & 0.15              & 0.14        & 0.14                    & 0.21      & 0.21                  & 0.10  & 0.10              & 0.44  & 0.44              \\ \midrule
    \goodflip            & 0     & 0                 & 0           & 0                       & 0         & 0                     & 0     & 0                 & 0     & 0                 \\ \toprule
    
    ~   & \textbf{AV-6}  & \textbf{\exescanner} & \textbf{AV-7}  & \textbf{\exescanner} & \textbf{AV-8}  & \textbf{\exescanner} & \textbf{AV-9}  & \textbf{\exescanner} & \textbf{AV-10} & \textbf{\exescanner} \\ \midrule
    Accuracy      & 91.25  & 95.91              & 93.18      & 97.05                  & 95.67    & 96.96                & 94.69 & 97.78             & 98.88     & 99.20                 \\
    Precision     & 99.27  & 99.36              & 99.47      & 99.53                  & 99.65    & 99.66                & 99.79 & 99.81             & 99.72     & 99.73                 \\
    Recall        & 78.86  & 90.42              & 83.52      & 93.13                  & 89.57    & 92.77                & 87.00 & 94.68             & 97.48     & 98.28                 \\
    Recall (unm.) & 99.27  & 99.27              & 97.98      & 97.98                  & 99.69    & 99.74                & 97.93 & 98.08             & 99.53     & 99.53                 \\
    Recall (adv.) & 76.12  & 89.23              & 81.58      & 92.48                  & 88.22    & 91.84                & 85.53 & 94.23             & 97.21     & 98.12                 \\
    F1            & 87.89  & 94.68              & 90.80      & 96.22                  & 94.34    & 96.09                & 92.96 & 97.18             & 98.59     & 99.00                 \\
    FPR           & 0.39   & 0.39               & 0.3        & 0.3                    & 0.21     & 0.21                 & 0.12  & 0.12              & 0.18      & 0.18                  \\ \midrule
    \goodflip     & 0      & 0                  & 0          & 0                      & 0        & 0                    & 0     & 0                 & 0         & 0                     \\ \bottomrule 
    \end{tabular}
    }
\end{table*}

The results are presented in \autoref{table:distributed_EXE-scanner_AVs_0.01}, where we report the performance metrics of AVs extended by \exescanner plugin with threshold setting to meet 0.01\% FPR (as described in \autoref{sec:transfer}).
We mimic the inclusion of \exescanner to the AVs by forwarding their response to our detector in case of benign prediction.

Hence, since this setting can be considered as an update to the original detection malware pipeline, and we do not re-analyze samples already flagged by AV as malicious, we only include the total number of goodware that are newly misclassified by our system (\goodflip).
This is in contrast with enhancing models using AT, which can cause an increase in both \goodflip and \malflip.
Aligned with findings of recent work~\citep{song2022mab, kozak2023creating, demetrio2021functionality}, we remark that some transfer attacks can be effective against commercial AVs.
Hence, once coupled to \exescanner, we clearly see that our detector is enhancing the robustness to adversarial EXEmples of AVs.
This can be appreciated by reading the ``Recall (Adv.)'' rows of \autoref{table:distributed_EXE-scanner_AVs_0.01} of the ``EXE-scanner'' column, where all but one AV now reach over 90\% of detection of adversarial EXEmples.

Similarly to what we have observed in \autoref{table:comparison_AE_detector_and_AT}, trained \exescanner is not causing any regression of benign predictions and no measurable FPR increase. However, we note that when AVs are paired with \exescanner with 1\% FPR threshold, there is a slight increase in FPR accompanied by four \goodflip{}s as shown in \autoref{table:distributed_EXE-scanner_AVs_1} in \autoref{appendix:A}.
Nevertheless, this regression is again minimal, never exceeding 0.02\% of false alarms, while improving detection rates of both unmodified malicious and adversarial EXEmples.
It follows that \exescanner can be safely distributed across diverse AV engines without the need for any changes to the plugin.

\subsection{Adaptive Attacks against \exescanner}
\label{sec:exescanner_robustness}
Next, we want to quantify the robustness of our methodology against adversarial EXEmples computed on the combination of the base classifier backed up by \exescanner.
This is achieved by running our generatos directly against \exescanner, thus leverigin its prediction while optimizing the creation of adversarial EXEmples.
We remark that the previous evaluations were considering transfer evaluations, where adversarial EXEmples were computed on other models and later tested on \exescanner.
Since these evasion attempts are suboptimal due to the fact that they have been computed without considering the predictions of our methodology, we now selected 500 unmodified malware samples from the $\text{test}_1$ set and seven black-box generators of adversarial EXEmples to create EXEmples against the original target classifiers enhanced by \exescanner. 
Due to the non-differentiable nature of our detector, we only use \emph{gradient-free} generators of adversarial EXEmples.
In particular, we operate with AMG-PPO, AMG-Random, GAMMA, SC-Gym-Malware, BB-Gym-Malware, MAB-EMBER, and MAB-MalConv.
All the strategies are set up in the same way as described in \autoref{sec:proposed_method_dataset}, except that now the target model to evade is the combination of the target classifier with \exescanner on top of it.
We show results of the robustness of our plugin detector in \autoref{table:EXE_scanner_under_attack}.

\begin{table}[h!]
    \centering
    \caption{Detection rates (recall) of adversarial EXEmples targeted against the original target classifier  (\textbf{G} stands for GBDT, and \textbf{M} stands for MalConv), and when extended with \exescanner. The threshold of \exescanner is set at 1\% FPR as described in \autoref{sec:transfer}.}
    \label{table:EXE_scanner_under_attack}
    \resizebox{\columnwidth}{!}{%
    \begin{tabular}{@{}lccc@{}}
        \toprule
        ~ & \textbf{Target Classifier} & \textbf{EXE-scanner} & \textbf{Increase} \\ \midrule
        AMG-PPO               & (G) 48.36    & 86.92       & 38.56     \\
        AMG-random            & (G) 66.43    & 87.14       & 20.71     \\
        GAMMA                 & (M) 59.23    & 99.06       & 39.83     \\
        SC-Gym-Malware        & (G) 93.51    & 98.17       & 4.66      \\
        BB-Gym-Malware        & (G) 93.51    & 98.64       & 5.13      \\
        MAB-MalConv           & (M) 10.45    & 17.92       & 7.47      \\
        MAB-EMBER             & (G) 29.44    & 38.72       & 9.28      \\ \bottomrule
    \end{tabular}
    }
\end{table}

The ``Target Classifier'' and ``EXE-scanner'' columns represent the detection rates of adversarial EXEmples generated against original target classifiers (see \autoref{sec:adversarial_EXEmples}) and the same classifiers extended by \exescanner, respectively.
Since we are considering a smaller subset of data, the detection rates by the original target classifier might be different from the ones reported in \autoref{table:patching_classifiers} where we used the entire $\text{test}_1$ set.
From the last column of \autoref{table:EXE_scanner_under_attack}, where we state the percentage increase of detection rate for \exescanner enhanced classifiers, we can see that it is harder for each of the seven generators to create evasive adversarial EXEmples against \exescanner extended classifiers.
Notably, AMG-PPO, AMG-random, and GAMMA, which were generating evasive EXEmples in 33\%-51\% of cases against the GBDT and MalConv detectors, decreased their performance down to 1\%-13\% against \exescanner enhanced detectors.

The MAB-EMBER and MAB-MalConv generators were the hardest to detect by the original target classifier, while enhancing the classifiers with \exescanner decreases the possibility of generating evasive EXEmples by 7\%-9\%, these generators still remain the hardest to protect against.

\subsection{Degradation over Time}
Previously, all experimental evaluation was done using $\text{test}_1$ set, a set of samples coming from the same time distribution as the train set. 
While it is valuable to measure the performance of malware detectors against contemporary threats, it is similarly important to periodically assess the performance to ensure appropriate updates of detectors. 
We mimic this scenario by utilizing the $\text{test}_2$ set described in \autoref{sec:proposed_method_dataset}. 
In the following experiments, we do not readjust the thresholds of tested models to specific FPR levels.
Rather, we use the thresholds previously set in \autoref{sec:transfer} and \autoref{sec:at_exescanner}, allowing us to evaluate the performance of models when not updated.
Hence, when we refer to evaluation at an explicit FPR level, we present the results on the $\text{test}_2$ with models using the original threshold computed from $\text{test}_1$.

We position \exescanner plugin as a direct competitor to adversarial training in the domain of protecting malware detectors against adversarial EXEmples. 
As such, we are interested in a comparison of how these protection methods degrade over time and how they perform when not updated.
We report the comparison results at 1\%, 0.1\%, and 0.01\% FPR levels on the $\text{test}_2$ set in \autoref{table:time_comparison_AE_detector_and_AT}.

\begin{table*}[h!]
    \centering
    \caption{Comparison of incremental and from scratch AT with \exescanner extended malware classifier on the $\text{test}_2$ set at 1\% (\ref{subtable:time_comparison_AE_detector_and_AT_1}), 0.1\% (\ref{subtable:time_comparison_AE_detector_and_AT_0.1}), and 0.01\% (\ref{subtable:time_comparison_AE_detector_and_AT_0.01}) thresholds. We report both \malflip and \goodflip regressions occurring while updating models.}
    \label{table:time_comparison_AE_detector_and_AT}
    \begin{subtable}{0.5\textwidth}
        \centering
        \resizebox{0.95\columnwidth}{!}{%
        \begin{tabular}{@{}lcccc@{}}
            \toprule
            ~ & \textbf{Baseline} & \textbf{\exescanner} & \textbf{AT-incremental} & \textbf{AT-full} \\ \midrule
            Accuracy      & 90.59    & 90.61       & 92.30          & 86.58   \\
            Precision     & 85.71    & 85.74       & 86.62          & 81.30   \\
            Recall        & 70.75    & 70.87       & 78.59          & 53.90   \\
            Recall (unm.) & 80.14    & 80.24       & 84.11          & 71.20   \\
            Recall (adv.) & 69.58    & 69.70       & 77.90          & 51.75   \\
            F1            & 77.52    & 77.60       & 82.41          & 64.83   \\ 
            FPR           & 3.51     & 3.51        & 3.61           & 3.69    \\ \midrule
            \goodflip     & 0        & 0           & 113            & 265     \\
            \malflip      & 0        & 0           & 45             & 1640    \\ \bottomrule
        \end{tabular}
        }
    \caption{1\% FPR}
    \label{subtable:time_comparison_AE_detector_and_AT_1}
    \end{subtable}%
    \begin{subtable}{0.5\textwidth}
        \centering
        \resizebox{0.95\columnwidth}{!}{%
        \begin{tabular}{@{}lcccc@{}}
            \toprule
            ~ & \textbf{Baseline} & \textbf{\exescanner} & \textbf{AT-incremental} & \textbf{AT-full} \\ \midrule
            Accuracy      & 88.63    & 88.79       & 88.32          & 78.73   \\
            Precision     & 99.78    & 99.79       & 99.40          & 95.96   \\
            Recall        & 50.54    & 51.23       & 49.37          & 7.57    \\
            Recall (unm.) & 75.87    & 75.97       & 75.77          & 5.96    \\
            Recall (adv.) & 47.39    & 48.14       & 46.07          & 7.78    \\
            F1            & 67.10    & 67.70       & 65.97          & 14.04   \\
            FPR           & 0.03     & 0.03        & 0.09           & 0.10    \\ \midrule
            \goodflip     & 0        & 0           & 18             & 21      \\
            \malflip      & 0        & 0           & 218            & 3975    \\ \bottomrule 
        \end{tabular}
        }
    \caption{0.1\% FPR}
    \label{subtable:time_comparison_AE_detector_and_AT_0.1}
    \end{subtable}%
    \\
    \begin{subtable}{0.5\textwidth}
        \centering
        \resizebox{0.95\columnwidth}{!}{%
        \begin{tabular}{@{}lcccc@{}}
            \toprule
            ~ & \textbf{Baseline} & \textbf{\exescanner} & \textbf{AT-incremental} & \textbf{AT-full} \\ \midrule
            Accuracy      & 83.69    & 84.23       & 82.10          & 77.38   \\
            Precision     & 100.00   & 100.00      & 99.90          & 97.06   \\
            Recall        & 28.89    & 31.23       & 21.99          & 1.45    \\
            Recall (unm.) & 50.55    & 52.14       & 39.62          & 0.99    \\
            Recall (adv.) & 26.19    & 28.63       & 19.79          & 1.51    \\
            F1            & 44.83    & 47.60       & 36.04          & 2.86    \\
            FPR           & 0.00     & 0.00        & 0.01           & 0.01    \\ \midrule
            \goodflip     & 0        & 0           & 2              & 4       \\
            \malflip      & 0        & 0           & 683            & 2520    \\ \bottomrule
        \end{tabular}
        }
    \caption{0.01\% FPR}
    \label{subtable:time_comparison_AE_detector_and_AT_0.01}
    \end{subtable}%
\end{table*}

At first glance, we can see a rapid decrease in the effectiveness of \exescanner and AT protection methods compared to testing on a dataset coming from the same time distribution as the training set presented in \autoref{sec:at_exescanner}.
In particular, AT-full under-performs the original baseline algorithm across all tested FPR levels. 
While AT-incremental provides the highest level of robustness against adversarial EXEmples and overall highest accuracy and F1 score at 1\% FPR, at lower levels, its performance is surpassed by both baseline and \exescanner extended detectors.
Furthermore, we recorded a significant increase in the goodware and malware flips by AT-full and AT-incremental over the already high regression counts from the previous testing on $\text{test}_1$ set (see \autoref{table:comparison_AE_detector_and_AT}).
This is in contrast with \exescanner prediction behavior, which does not introduce any prediction regressions across all tested FPR levels and both test sets.

Similarly to our experiments in \autoref{sec:at_exescanner}, we repeat the 10:1 experiment on test$_2$, and we report all the results in \autoref{table:time_comparison_AE_detector_and_AT_10:1} in \autoref{appendix:B}.
Our findings are aligned with those on \autoref{table:time_comparison_AE_detector_and_AT}, confirming once more the effectiveness of \exescanner.

\subsection{Summary of Results}
The experimental analysis revealed several key findings about the effectiveness of \exescanner in mitigating adversarial EXEmples:
(i), the majority of adversarial EXEmples that are optimized against target classifiers can be successfully halted by extending the target classifiers with \exescanner.
This demonstrates the efficiency of trained \exescanner in dealing with known attacks; 
(ii), while transfer attacks are generally less potent than direct attacks, \exescanner still significantly improves the robustness of the attacked models. 
This underscores the broad range of possible applications for \exescanner;
(iii), our a-posteriori analysis of the behavior of the predictions of \exescanner{} reveals feature space artifacts left behind by the generators of adversarial EXEmples, allowing models to detect their perturbations;
(iv), \exescanner is a potential alternative to adversarial training. 
It yields improvements comparable or better to from scratch and incremental adversarial training scenarios and causes significantly less regression on previously correctly labeled samples than either of them. 
This suggests that \exescanner could be a more efficient method for strengthening the robustness of malware classifiers;
(v), even commercially-available AV software hosted on VirusTotal is susceptible to adversarial EXEmples.
However, when top-tier AV is paired with \exescanner, this vulnerability can be mitigated while maintaining a low false positive rate;
(vi), while feasible, generating new adversarial EXEmples against models enhanced with \exescanner is substantially more challenging than before. This further attests to the robustness of \exescanner in defending against EXEmples;
(vii), the performance of adversarial training and \exescanner degrades over time when models are not updated.
However, even on unseen data from different time distribution, \exescanner plugin provides more efficient protection against EXEmples with decreasing levels of FPR while maintain not incurring regressions to predictions of the original classifier. 

In conclusion, the findings from the experimental analysis strongly suggest that \exescanner is an effective tool for strengthening malware classifiers' resilience against adversarial EXEmples.
\section{Related Work} \label{sec:related_work}
In this paper, we discussed the inclusion of \exescanner as a plugin to the malware detection pipeline to stop the spread of adversarial EXEmples.
However, there are other techniques that can be placed inside pipelines to withstand attacks using adversarial EXEmples against Windows malware detection.

\mypar{Detection of Adversarial EXEmples}
Alasmary et al.~\citep{alasmary2020soteria} propose a technique for detecting adversarial EXEmples by analyzing the Control Flow Graph (CFG) of input samples.
These graphs represent the structure of the program, with all its loops and branches.
From CFGs, they then proceed by exploring it through random walks, later processed to be fed to an autoencoder model as training data.
Hence, detection is achieved by comparing the error of the original walk in the graph and its reconstruction.
If such error is higher than a threshold, the sample is flagged as adversarial. Non-adversarial samples are processed by CNN classifier for final classification.
While this method looks promising, our \exescanner is easier to test and deploy, since it does not rely on extracting any CFG from samples, and it is trained on a well-known and already-available feature representation.
Also, this technique was evaluated on a limited range of adversarial perturbations, thus it is unclear how well it generalizes to other types of adversarial EXEmples.
Lastly, no code is available to either replicate their results or compare them with \exescanner.

\mypar{Robust Models against Adversarial EXEmples.}
Quiring et al.~\citep{quiring2020against} propose a detector that strips all the so-called \emph{semantic gaps}, which refer to all the commonly-unused space inside executables where adversarial content might be placed to fool detection.
On top of this, they leverage an ensemble of models, ranging from static signatures to neural networks and GBDTs.
Also, they include a stateful defense that computes similarities between analyzed samples to understand if there is an ongoing attack.
Similarly, Abusnaina et al.~\citep{abusnaina2023burning} propose a technique that first applies pre-processing steps to eliminate and nullify the effect of manipulations that fill unused space inside binaries, like removing the padding, reset to zeros the content between sections, and the removal of unnecessary information from PE format.
To face the injection of sections, the authors used graph-based representation for binaries, where nodes consist of individual sections of the original PE file represented as n-grams, bytes histograms and extracted strings.
Both methodologies rely on pre-processing steps and multiple models, while \exescanner just needs one feature extraction step and one forward operation.
Also, these models require re-training the deployed model from scratch, without analyzing the regression that occurs during such an update.

\mypar{Certified Robustness.}
Huang et al.~\citep{huang2023rs} propose RS-Del, a certification approach that statistically guarantees the inexistence of adversarial EXEmples within a perturbation radius, defined through the Levenshtein distance.
RS-Del uses a baseline MalConv model trained on small portions of executables computed by discarding more than 90\% of the original content.
Then, at test time, RS-Del samples 1000 chunks of bytes from the input program similarly to what is done at training time, and then it proceeds by computing a prediction through majority voting.
Similarly, Gibert et al.~\citep{gibert2023certified} divide the input programs into chunks of contiguous bytes that are used to train a MalConv model.
Hence, at test time, they again divide the sample into different chunks, and prediction is computed through majority voting.
While both techniques are timely since they are the first major works that certify the robustness of malware detection, they require extensive and resource-demanding training processes.
The majority voting mechanism requires houndreads for predictions since samples are divided in chunks indipendently passed in input to the model, and they still suffer against content injection attacks that greatly enlarge the filesize.
On the contrary, \exescanner is trained only once on data, and it requires only one prediction after the baseline model.
Also, from our results, we have shown that \exescanner can detect and withstand large content injection attacks.
Lastly, these models require re-training the deployed model from scratch, without analyzing the regression that incurr during such an update.


\section{Limitations, Future Work and Conclusions}
\label{sec:conclusion}
We conclude our work by discussing the limitations of our methodology, along with possible future lines of research based on our findings.

\mypar{Limitations.} 
Even though we provide an extensive experimental analysis, this study has several limitations that should be considered. Firstly, the architecture of \exescanner was confined to a GBDT model with pre-defined hyperparameters.
However, while this lack of model diversity may limit the potential of our technique, we also showed in \autoref{sec:evaluation_exe_scanner_architectures} that the GBDT classifier has the best trade-off in terms of performance w.r.t. the other competitors.
Secondly, the training malware dataset was sourced exclusively from the VirusShare repository and was found to contain predominantly Trojan malware samples.
This limited source of data may introduce a bias in our results and restrict the range of adversarial EXEmples our plugin can effectively identify. 
Lastly, the \exescanner was subjected to only a selected set of generators (AMG-PPO, AMG-Random, GAMMA, SC-Gym-Malware, BB-Gym-Malware, MAB-MalConv, MAB-EMBER). 
Therefore, its robustness against other potential generators of adversarial EXEmples remains untested.

\mypar{Future Work.} 
Future work should aim to address the aforementioned limitations by extending the model selection process for \exescanner, while also expanding both the sources of goodware and malware samples, both selected from other time intervals and validated through the lens of VirusTotal.
We plan to improve the testing of \exescanner against a broader range of adversarial EXEmples, by enlarging the experiments with adversarial EXEmples created by behavioral manipulations as proposed by Lucas et al.~\citep{lucas2023adversarial}. 
Including more varieties of adversarial perturbations could help create a more effective and robust solution. 
Next, we intend to explore the area of dynamic malware analysis and how dynamic features (e.g., behavior logs) could be used for mitigating adversarial EXEmples.
We anticipate that this direction could lead to significant changes in how we view the EXEmples of today.

\mypar{Conclusions.}
We propose \exescanner, a lightweight plugin designed to increase the robustness of malware detectors against adversarial EXEmples. 
With extensive experimental analysis, we show that the inclusion of \exescanner increases the detection of adversarial EXEmples up to 99\%, thus preventing adversarial EXEmples from being effective against the defended model.
Also, \exescanner achieves a similar increase in robustness as adversarial training, which, on the contrary, causes regression of performance by misclassifying previously correctly predicted samples.
To exacerbate this setting, we also quantify the degradation caused by the drifts in the data distribution, highlighting that, thanks to \exescanner, the base classifier is less prone to a regression of accuracy rather than its competitors.
We extend our analysis also to commercially-available products, highlighting that \exescanner can be coupled to AVs without any additional training or fine-tuning, increasing their robustness with only a negligible increment of FPR.
Through the lens of SHAP, we highlight that adversarial EXEmples contain artifacts that depend on the applied manipulation, giving \exescanner the ability to detect them with ease.
Lastly, we show that \exescanner is also harder to attack when targeted by specific adversarial EXEmples crafted on its responses, thanks to both the usage of only binary responses and its robust training.
Thanks to its lightweight design and easy usage, cybersecurity engineers can swiftly deliver security updates against rapidly evolving adversarial EXEmples by embracing our methodology with minimal effort.

\mypar{Final Remarks.}
To promote future work and reproducibility of our results, we publish source codes of \exescanner and collected datasets of goodware, unmodified malware, and adversarial EXEmples for the cybersecurity community. 

\section*{Acknowledgments}
This work was supported by MEYS of the Czech Republic, grant No. SGS23/211/OHK3/3T/18 of the Grant Agency, Czech Technical University in Prague; and SERICS (PE00000014) under the MUR National Recovery and Resilience Plan funded by the European Union – NextGenerationEU.
\section*{Availability}
We share the code of \exescanner and data used for training and evaluation at the following address \url{https://github.com/matouskozak/EXE-scanner}.


\bio{}
\bibliographystyle{cas-model2-names.bst}
\bibliography{main}
\endbio

\appendix
\section{Appendix}
\label{appendix:A}
We include here all the results of \exescanner when coupled with different commercial AVs.

\begin{table*}[h!]
    \centering
    \caption{Performance of \exescanner (1\% FPR threshold) with different AV engines on $\text{test}_1$ set.}
    \label{table:distributed_EXE-scanner_AVs_1}
    \resizebox{\textwidth}{!}{%
    \begin{tabular}{@{}l|cc|cc|cc|cc|cc@{}}
    \toprule
    ~ & \textbf{AV-1}  & \textbf{\exescanner} & \textbf{AV-2}  & \textbf{\exescanner} & \textbf{AV-3}  & \textbf{\exescanner} & \textbf{AV-4}  & \textbf{\exescanner} & \textbf{AV-5}  & \textbf{\exescanner} \\ \midrule
    Accuracy      & 94.18 & 98.81             & 94.68       & 98.76                   & 92.59     & 98.10                 & 93.95 & 97.90             & 94.96 & 98.72             \\
    Precision     & 99.74 & 99.74             & 99.77       & 99.77                   & 99.63     & 99.66                 & 99.83 & 99.83             & 99.26 & 99.31             \\
    Recall        & 85.77 & 97.29             & 86.99       & 97.14                   & 81.91     & 95.61                 & 85.13 & 94.96             & 88.15 & 97.49             \\
    Recall (unm.) & 98.91 & 99.12             & 97.88       & 98.24                   & 96.32     & 96.63                 & 94.87 & 95.34             & 98.13 & 98.45             \\
    Recall (adv.) & 84.01 & 97.05             & 85.53       & 96.99                   & 79.98     & 95.48                 & 83.83 & 94.91             & 86.81 & 97.37             \\
    F1            & 92.23 & 98.50             & 92.94       & 98.44                   & 89.90     & 97.59                 & 91.90 & 97.33             & 93.37 & 98.39             \\
    FPR           & 0.15  & 0.17              & 0.14        & 0.15                    & 0.21      & 0.22                  & 0.10  & 0.11              & 0.44  & 0.46              \\ \midrule
    \goodflip     & 0     & 4                 & 0           & 4                       & 0         & 4                     & 0     & 4                 & 0     & 4                 \\ \toprule
    
    ~   & \textbf{AV-6}  & \textbf{\exescanner} & \textbf{AV-7}  & \textbf{\exescanner} & \textbf{AV-8}  & \textbf{\exescanner} & \textbf{AV-9}  & \textbf{\exescanner} & \textbf{AV-10} & \textbf{\exescanner} \\ \midrule
    Accuracy      & 91.25  & 98.00              & 93.18      & 98.31                  & 95.67    & 98.49                & 94.69 & 98.80             & 98.88     & 99.32             \\
    Precision     & 99.27  & 99.37              & 99.47      & 99.52                  & 99.65    & 99.65                & 99.79 & 99.79             & 99.72     & 99.70             \\
    Recall        & 78.86  & 95.63              & 83.52      & 96.26                  & 89.57    & 96.58                & 87.00 & 97.23             & 97.48     & 98.61             \\
    Recall (unm.) & 99.27  & 99.27              & 97.98      & 98.03                  & 99.69    & 99.79                & 97.93 & 98.34             & 99.53     & 99.64             \\
    Recall (adv.) & 76.12  & 95.14              & 81.58      & 96.03                  & 88.22    & 96.15                & 85.53 & 97.08             & 97.21     & 98.47             \\
    F1            & 87.89  & 97.47              & 90.80      & 97.86                  & 94.34    & 98.09                & 92.96 & 98.49             & 98.59     & 99.15             \\
    FPR           & 0.39   & 0.41               & 0.3        & 0.31                   & 0.21     & 0.23                 & 0.12  & 0.14              & 0.18      & 0.20              \\ \midrule
    \goodflip     & 0      & 4                  & 0          & 4                      & 0        & 4                    & 0     & 4                 & 0         & 4                \\ \bottomrule 
    \end{tabular}
    }
\end{table*}

\begin{table*}[h!]
    \centering
    \caption{Performance of \exescanner (0.1\% FPR threshold) with different AV engines on $\text{test}_1$ set.}
    \label{table:distributed_EXE-scanner_AVs_0.1}
    \resizebox{\textwidth}{!}{%
    \begin{tabular}{@{}l|cc|cc|cc|cc|cc@{}}
    \toprule
    ~ & \textbf{AV-1}  & \textbf{\exescanner} & \textbf{AV-2}  & \textbf{\exescanner} & \textbf{AV-3}  & \textbf{\exescanner} & \textbf{AV-4}  & \textbf{\exescanner} & \textbf{AV-5}  & \textbf{\exescanner} \\ \midrule
Accuracy      & 94.18 & 97.65             & 94.68       & 97.73                   & 92.59     & 96.75                 & 93.95 & 96.63             & 94.96 & 97.81             \\
Precision     & 99.74 & 99.76             & 99.77       & 99.79                   & 99.63     & 99.67                 & 99.83 & 99.85             & 99.26 & 99.32             \\
Recall        & 85.77 & 94.39             & 86.99       & 94.58                   & 81.91     & 92.23                 & 85.13 & 91.77             & 88.15 & 95.21             \\
Recall (unm.) & 98.91 & 98.96             & 97.88       & 97.98                   & 96.32     & 96.37                 & 94.87 & 94.97             & 98.13 & 98.24             \\
Recall (adv.) & 84.01 & 93.77             & 85.53       & 94.12                   & 79.98     & 91.68                 & 83.83 & 91.34             & 86.81 & 94.80             \\
F1            & 92.23 & 97.00             & 92.94       & 97.11                   & 89.90     & 95.81                 & 91.90 & 95.64             & 93.37 & 97.22             \\
FPR           & 0.15  & 0.15              & 0.14        & 0.14                    & 0.21      & 0.21                  & 0.10  & 0.10              & 0.44  & 0.44              \\ \midrule
\goodflip     & 0     & 0                 & 0           & 0                       & 0         & 0                     & 0     & 0                 & 0     & 0                 \\ \toprule
    
    ~   & \textbf{AV-6}  & \textbf{\exescanner} & \textbf{AV-7}  & \textbf{\exescanner} & \textbf{AV-8}  & \textbf{\exescanner} & \textbf{AV-9}  & \textbf{\exescanner} & \textbf{AV-10} & \textbf{\exescanner} \\ \midrule
Accuracy      & 91.25  & 95.91              & 93.18      & 97.05                  & 95.67    & 96.96                & 94.69 & 97.78             & 98.88     & 99.20                 \\
Precision     & 99.27  & 99.36              & 99.47      & 99.53                  & 99.65    & 99.66                & 99.79 & 99.81             & 99.72     & 99.73                 \\
Recall        & 78.86  & 90.42              & 83.52      & 93.13                  & 89.57    & 92.77                & 87.00 & 94.68             & 97.48     & 98.28                 \\
Recall (unm.) & 99.27  & 99.27              & 97.98      & 97.98                  & 99.69    & 99.74                & 97.93 & 98.08             & 99.53     & 99.53                 \\
Recall (adv.) & 76.12  & 89.23              & 81.58      & 92.48                  & 88.22    & 91.84                & 85.53 & 94.23             & 97.21     & 98.12                 \\
F1            & 87.89  & 94.68              & 90.80      & 96.22                  & 94.34    & 96.09                & 92.96 & 97.18             & 98.59     & 99.00                 \\
FPR           & 0.39   & 0.39               & 0.30       & 0.30                   & 0.21     & 0.21                 & 0.12  & 0.12              & 0.18      & 0.18                  \\ \midrule
\goodflip     & 0      & 0                  & 0          & 0                      & 0        & 0                    & 0     & 0                 & 0         & 0                     \\ \bottomrule 
    \end{tabular}
    }
\end{table*}

\newpage

\section{Appendix}
\label{appendix:B}
We place here all the results obtained by considering an highly-imbalanced scenario of 10:1 ratio between goodware and malware samples.

\begin{table*}[h!]
    \centering
    \caption{
    Comparison of incremental and from scratch AT with \exescanner extended malware classifier on the $\text{test}_1$ where we impose an imbalance of 10:1 goodware to malware.
    Results are shown at 1\% (\ref{subtable:comparison_AE_detector_and_AT_10:1_1}), 0.1\% (\ref{subtable:comparison_AE_detector_and_AT_10:1_0.1}), and 0.01\% (\ref{subtable:comparison_AE_detector_and_AT_10:1_0.01}) FPR thresholds.
    We report both \malflip and \goodflip regressions occurring while updating models.}
    \label{table:comparison_AE_detector_and_AT_10:1}
    \begin{subtable}{0.5\textwidth}
        \centering
        \resizebox{0.95\columnwidth}{!}{%
        \begin{tabular}{@{}lcccc@{}}
            \toprule
            ~ & \textbf{Baseline} & \textbf{\exescanner} & \textbf{AT-incremental} & \textbf{AT-full} \\ \midrule
Accuracy      & 98.30    & 98.79       & 98.85          & 98.94   \\
Precision     & 90.13    & 90.63       & 90.72          & 90.86   \\
Recall        & 91.32    & 96.69       & 97.36          & 98.18   \\
Recall (unm.) & 89.29    & 90.87       & 92.46          & 94.44   \\
Recall (adv.) & 91.56    & 97.37       & 97.92          & 98.62   \\ 
F1            & 90.72    & 93.56       & 93.92          & 94.38   \\ \midrule
\goodflip     & 0        & 0           & 56             & 154     \\
\malflip      & 0        & 0           & 4              & 18      \\ \bottomrule
        \end{tabular}
        }
    \caption{1\% FPR}
    \label{subtable:comparison_AE_detector_and_AT_10:1_1}
    \end{subtable}%
    \begin{subtable}{0.5\textwidth}
        \centering
        \resizebox{0.95\columnwidth}{!}{%
        \begin{tabular}{@{}lcccc@{}}
            \toprule
            ~ & \textbf{Baseline} & \textbf{\exescanner} & \textbf{AT-incremental} & \textbf{AT-full} \\ \midrule
Accuracy      & 95.67    & 97.86       & 96.84          & 98.68   \\
Precision     & 98.18    & 98.73       & 98.52          & 98.87   \\
Recall        & 53.39    & 77.40       & 66.20          & 86.49   \\
Recall (unm.) & 57.14    & 63.49       & 71.43          & 78.57   \\
Recall (adv.) & 52.95    & 79.01       & 65.59          & 87.41   \\
F1            & 69.16    & 86.77       & 79.19          & 92.26   \\ \midrule
\goodflip     & 0        & 0           & 8              & 16      \\
\malflip      & 0        & 0           & 11             & 101     \\ \bottomrule
        \end{tabular}
        }
    \caption{0.1\% FPR}
    \label{subtable:comparison_AE_detector_and_AT_10:1_0.1}
    \end{subtable}%
    \\
    \begin{subtable}{0.5\textwidth}
        \centering
        \resizebox{0.95\columnwidth}{!}{%
        \begin{tabular}{@{}lcccc@{}}
            \toprule
            ~ & \textbf{Baseline} & \textbf{\exescanner} & \textbf{AT-incremental} & \textbf{AT-full} \\ \midrule
            Accuracy      & 93.98    & 97.28       & 94.55          & 93.79   \\
Precision     & 99.76    & 99.88       & 99.79          & 99.74   \\
Recall        & 33.88    & 70.21       & 40.12          & 31.82   \\
Recall (unm.) & 42.86    & 50.40       & 45.63          & 24.60   \\
Recall (adv.) & 32.84    & 72.51       & 39.48          & 32.66   \\
F1            & 50.59    & 82.46       & 57.24          & 48.25   \\ \midrule
\goodflip     & 0        & 0           & 1              & 2       \\
\malflip      & 0        & 0           & 23             & 353     \\ \bottomrule
        \end{tabular}
        }
    \caption{0.01\% FPR}
    \label{subtable:comparison_AE_detector_and_AT_10:1_0.01}
    \end{subtable}%
\end{table*}

\begin{table*}[h!]
    \centering
    \caption{Comparison of incremental and from scratch AT with \exescanner extended malware classifier on the $\text{test}_2$ where we impose an imbalance of 10:1 goodware to malware. Results are shown at 1\% (\ref{subtable:time_comparison_AE_detector_and_AT_10:1_1}), 0.1\% (\ref{subtable:time_comparison_AE_detector_and_AT_10:1_0.1}), and 0.01\% (\ref{subtable:time_comparison_AE_detector_and_AT_10:1_0.01}) FPR thresholds. We report both \malflip and \goodflip regressions occurring while updating models.}
    \label{table:time_comparison_AE_detector_and_AT_10:1}
    \begin{subtable}{0.5\textwidth}
        \centering
        \resizebox{0.95\columnwidth}{!}{%
        \begin{tabular}{@{}lcccc@{}}
            \toprule
            ~ & \textbf{Baseline} & \textbf{\exescanner} & \textbf{AT-incremental} & \textbf{AT-full} \\ \midrule
Accuracy      & 94.11    & 94.11       & 94.74          & 92.50   \\
Precision     & 66.69    & 66.71       & 68.40          & 59.55   \\
Recall        & 70.27    & 70.34       & 78.27          & 54.34   \\
Recall (unm.) & 81.07    & 81.07       & 84.53          & 72.27   \\
Recall (adv.) & 68.76    & 68.83       & 77.39          & 51.83   \\
F1            & 68.43    & 68.47       & 73.01          & 56.83   \\
FPR           & 3.51     & 3.51        & 3.61           & 3.69    \\ \midrule
\goodflip     & 0        & 0           & 113            & 265     \\
\malflip      & 0        & 0           & 9              & 524     \\ \bottomrule
        \end{tabular}
        }
    \caption{1\% FPR}
    \label{subtable:time_comparison_AE_detector_and_AT_10:1_1}
    \end{subtable}%
    \begin{subtable}{0.5\textwidth}
        \centering
        \resizebox{0.95\columnwidth}{!}{%
        \begin{tabular}{@{}lcccc@{}}
            \toprule
            ~ & \textbf{Baseline} & \textbf{\exescanner} & \textbf{AT-incremental} & \textbf{AT-full} \\ \midrule
Accuracy      & 95.44    & 95.52       & 95.32          & 91.53   \\
Precision     & 99.35    & 99.36       & 98.24          & 89.10   \\
Recall        & 50.21    & 51.00       & 49.36          & 7.77    \\
Recall (unm.) & 77.33    & 77.33       & 77.07          & 5.87    \\
Recall (adv.) & 46.41    & 47.31       & 45.48          & 8.03    \\
F1            & 66.71    & 67.40       & 65.71          & 14.29   \\
FPR           & 0.03     & 0.03        & 0.09           & 0.10     \\ \midrule
\goodflip     & 0        & 0           & 18             & 21      \\
\malflip      & 0        & 0           & 62             & 1321    \\ \bottomrule
        \end{tabular}
        }
    \caption{0.1\% FPR}
    \label{subtable:time_comparison_AE_detector_and_AT_10:1_0.1}
    \end{subtable}%
    \\
    \begin{subtable}{0.5\textwidth}
        \centering
        \resizebox{0.95\columnwidth}{!}{%
        \begin{tabular}{@{}lcccc@{}}
            \toprule
            ~ & \textbf{Baseline} & \textbf{\exescanner} & \textbf{AT-incremental} & \textbf{AT-full} \\ \midrule
Accuracy      & 93.59    & 93.82       & 92.96          & 91.06   \\
Precision     & 100.0    & 100.0       & 99.71          & 93.10   \\
Recall        & 29.50    & 32.02       & 22.62          & 1.77    \\
Recall (unm.) & 49.07    & 50.40       & 36.27          & 1.07    \\
Recall (adv.) & 26.76    & 29.45       & 20.70          & 1.87    \\
F1            & 45.56    & 48.51       & 36.87          & 3.47    \\
FPR           & 0.0      & 0.0         & 0.01           & 0.01    \\ \midrule
\goodflip     & 0        & 0           & 2              & 4       \\
\malflip      & 0        & 0           & 230            & 857     \\ \bottomrule
        \end{tabular}
        }
    \caption{0.01\% FPR}
    \label{subtable:time_comparison_AE_detector_and_AT_10:1_0.01}
    \end{subtable}%
\end{table*}

\end{document}